\documentclass[twocolumn,floatfix,showpacs,rmp,aps]{revtex4}
\pdfoutput=1
\usepackage{amssymb}
\usepackage{amsmath}
\usepackage{bm}
\usepackage[pdftex]{graphicx}

\begin{document}
\title{Topological Insulators}

\author{M. Z. Hasan}
\email{mzhasan@princeton.edu}
\affiliation{Joseph Henry Laboratories, Department of Physics,
Princeton University, Princeton, NJ 08544}

\author{C. L. Kane}
\email{kane@physics.upenn.edu}
\affiliation{Department of Physics and Astronomy,
University of Pennsylvania, Philadelphia, PA 19104}

\begin{abstract}
Topological insulators are electronic materials that
have a bulk band gap like an ordinary insulator, but have protected conducting
states on their edge or surface.  These states are possible due to the combination of spin orbit
interactions and time reversal symmetry.  The 2D topological insulator is a
quantum spin Hall insulator, which is a close cousin of the integer
quantum Hall state. A 3D topological
insulator supports novel spin polarized 2D Dirac fermions on its surface.  In this
Colloquium article we will review the theoretical foundation for
topological insulators and superconductors and describe recent experiments in which
the signatures of topological insulators have been observed.  We will describe transport
experiments on HgTe/CdTe quantum wells that demonstrate the
existence of the edge states predicted for the quantum spin Hall
insulator. We will then discuss experiments on Bi$_{1-x}$Sb$_x$, Bi$_2$Se$_3$,
Bi$_2$Te$_3$ and Sb$_2$Te$_3$ that establish these materials as 3D topological
insulators and directly probe the topology of their surface states.
We will then describe exotic
states that can occur at the surface of a 3D topological
insulator due to an induced energy gap.  A magnetic gap leads to a
novel quantum Hall state that gives rise to a topological
magnetoelectric effect.  A superconducting energy gap leads to a
state that supports Majorana fermions, and may provide a new venue
for realizing proposals for topological quantum computation.  We will
close by discussing prospects for observing these exotic states, as
well as other potential device applications of topological
insulators.
\end{abstract}

\pacs{73.20.-r, 73.43.-f, 85.75.-d, 74.90.+n}
\maketitle
\tableofcontents

\section{Introduction}
\label{sec:intro}

A recurring theme in condensed matter physics
has been the discovery and classification of distinctive phases of matter.
Often, phases can be understood using Landau's approach, which characterizes
states in terms of underlying symmetries that are spontaneously
broken.   Over the past 30 years, the study
of the quantum Hall effect has led to a different classification paradigm, based
on the notion of {\it topological order} \cite{thouless82,wen95}.  The
state responsible for the quantum Hall effect does not
break any symmetries, but it defines a topological phase in the
sense that certain fundamental properties (such as the quantized value of the
Hall conductance, and the number of gapless boundary modes)
are insensitive to smooth changes in materials
parameters and can not change unless the system passes
through a quantum phase transition.

In the past five years a new field has emerged in condensed matter
physics, based on the realization that the spin orbit interaction can
lead to topological insulating electronic phases
\cite{kanemele05a,kanemele05b,moorebalents07,fukanemele07,roy09b}, and on the prediction
and observation of these phases in real materials
\cite{bernevighugheszhang06,fukane07,konig07,hsieh08,xia09a,zhangh09}.  A
topological insulator, like an ordinary insulator, has a bulk energy
gap separating the highest occupied electronic band from the lowest
empty band. The surface (or edge in two dimensions) of a topological
insulator, however, necessarily has gapless states that are
protected by time reversal symmetry.  The
topological insulator is closely related to the two dimensional (2D)
integer quantum Hall state, which also has unique
edge states.   The surface (or edge) states of a topological insulator lead
to a conducting state with properties unlike any other known
1D or 2D electronic systems.  In addition to their
fundamental interest, these states are predicted to have special
properties that could be useful for applications ranging from
spintronics to quantum computation.

The concept of topological order \cite{wen95} is often used to characterize the
intricately correlated fractional quantum Hall states \cite{tsui82}, which require an
inherently many body approach to understand \cite{laughlin83}.  However,
topological considerations also apply to the simpler
integer quantum Hall states \cite{thouless82}, for
which an adequate description can be formulated in terms of single
particle quantum mechanics.  In this regard, topological insulators
are similar to the integer quantum Hall effect.  Due to the presence of a
single particle energy gap, electron-electron
interactions do not modify the state in an essential way.
Topological insulators can be understood within the framework of the band
theory of solids \cite{bloch29}.  It is remarkable that after more than 80 years,
there are still treasures to be uncovered within band theory.

In this colloquium article we will review the theoretical and
experimental foundations of this rapidly developing field.  We
begin in Section \ref{sec:topobandtheory} with an introduction to
topological band theory, in which we will explain the topological
order in the quantum Hall effect and in topological insulators.
We will also give a short introduction to topological superconductors,
which can be understood within a similar framework.
A unifying feature of these states is the {\it bulk-boundary correspondence}, which
relates the topological structure of bulk crystal to the presence of gapless boundary modes.
Section III will describe the 2D topological insulator,
also known as a quantum spin Hall insulator and discuss the
discovery of this phase in HgCdTe quantum wells.
Section IV is devoted to 3D topological insulators.
We will review their experimental discovery in Bi$_{1-x}$Sb$_x$, as well as more
recent work on ``second generation" materials Bi$_2$Se$_3$ and
Bi$_2$Te$_3$.
Section V
will focus on exotic states that can occur at the surface of a topological
insulator due to an induced energy gap.  An energy gap induced by a magnetic
field or proximity to a magnetic material leads to a
novel quantum Hall state, along with a topological
magnetoelectric effect.  An energy gap due to proximity with a superconductor leads to a
state that supports Majorana fermions, and may provide a new venue
for realizing proposals for topological quantum computation.
In Section VI we will conclude with a discussion of new materials, new
experiments and open problems.

Some aspects of this subject have been described in other reviews, including
the review of the quantum spin Hall effect by \textcite{konig08} and surveys
by \textcite{qizhang10} and \textcite{moore10}.

\section{Topological Band Theory}
\label{sec:topobandtheory}

\subsection{The insulating state}
\label{sec:insulator}

The insulating state is the most basic state of matter.  The simplest insulator is
an atomic insulator, with electrons bound to atoms in closed shells.
Such a material is electrically inert because it takes a finite energy to
dislodge an electron.  Stronger interaction between atoms in a crystal
leads to covalent bonding.  One of the triumphs of quantum
mechanics in the 20th century was the development of the band theory of solids,
which provides a language for describing the electronic structure of such
states.  This theory exploits the translational symmetry of the
crystal to classify electronic states in terms of their crystal momentum
${\bf k}$, defined in a periodic Brillouin zone.  The Bloch states
$|u_m({\bf k})\rangle$, defined in a single unit cell of the crystal,
are eigenstates of the Bloch Hamiltonian ${\cal H}({\bf k})$.
The eigenvalues $E_m({\bf k})$
define energy bands that collectively form the band structure.
In an insulator an energy
gap separates the occupied valence band states from the empty
conduction band states.  Though the gap in an atomic
insulator, like solid Argon, is much larger than that of a
semiconductor, there is a sense in
which both belong to the same phase.  One can imagine
tuning the Hamiltonian so as to interpolate continuously between the
two without closing the energy gap.  Such a process defines a topological
equivalence between different insulating states.
If one adopts a slightly coarser
``stable" topological classification scheme, which
equates states with different numbers of trivial core bands,
then all conventional
insulators are equivalent.  Indeed, such insulators are
equivalent to the {\it vacuum}, which according to Dirac's
relativistic quantum theory also has an energy gap (for pair production),
a conduction band (electrons) and a valence band (positrons).

Are {\it all} electronic states with an
energy gap topologically equivalent to the vacuum?  The answer
is no, and the counterexamples are fascinating
states of matter.

\subsection{The quantum Hall state}
\label{sec:qhall}

\begin{figure}
\includegraphics[width=3in]{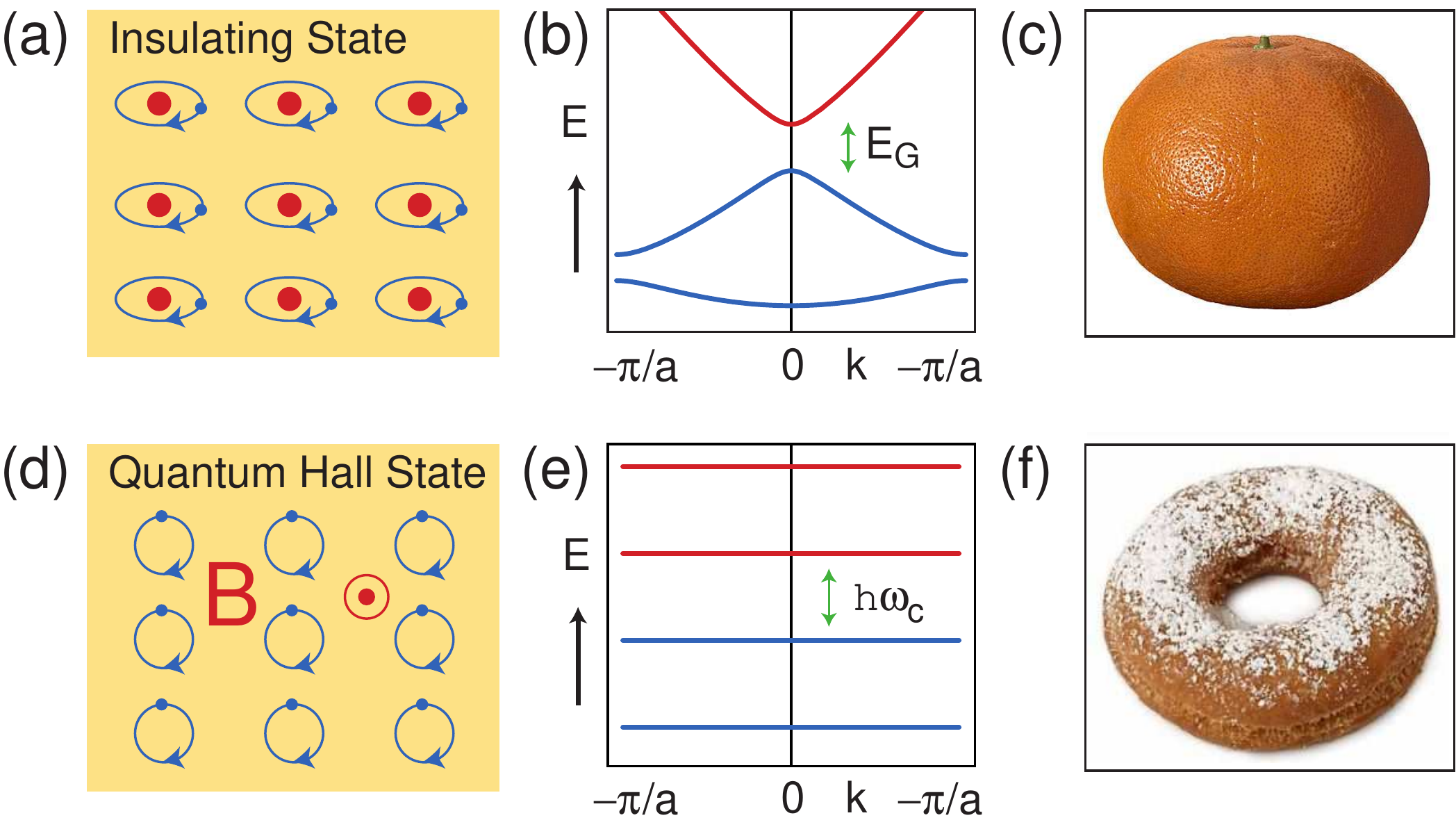}
\caption{(a, b, c) The insulating state.  (a) depicts an atomic insulator, while
(b) shows a simple model insulating band structure.  (d, e, f) The quantum Hall state.
(d) depicts the cyclotron motion of electrons, and (e) shows the Landau levels, which may
be viewed as a band structure.  (c) and (f) show two surfaces which differ in their genus, $g$.
$g=0$ for the sphere (c) and $g=1$ for the donut (f).  The Chern number $n$ that distinguishes
the two states is a topological invariant similar to the genus.}
\label{fig:insulator}
\end{figure}

The simplest counterexample is the integer quantum Hall state
\cite{vonklitzing80,prange87}, which
occurs when electrons confined to two dimensions are placed in a
strong magnetic field.  The quantization of the electrons' circular
orbits with cyclotron frequency $\omega_c$
leads to quantized Landau levels with
energy $\epsilon_m=\hbar\omega_c (m+1/2)$.
If $N$ Landau levels are filled and the rest are empty, then an energy gap
separates the occupied and empty states just like in an
insulator.  Unlike an insulator, though, an electric field causes
the cyclotron orbits to drift, leading to a Hall
current characterized by the quantized  Hall conductivity
\begin{equation}
\sigma_{xy} = N e^2/h.
\label{qhall}
\end{equation}
The quantization of $\sigma_{xy}$ has been measured to
one part in $10^9$ \cite{vonklitzing05}.  This precision is a
manifestation of the topological nature of $\sigma_{xy}$.

Landau levels can be viewed as a ``band
structure".  Since the generators of translations do not commute with one another
in a magnetic field, electronic states can
not be labeled with momentum.  However, if a unit cell with
area $2\pi \hbar c/e B$ enclosing a
flux quantum is defined, then {\it lattice} translations
{\it do} commute, so Bloch's theorem allows states to be labeled by
2D crystal momentum ${\bf k}$.  In the absence of a periodic potential,
the energy levels are simply the ${\bf k}$ independent
Landau levels, $E_m({\bf k}) = \epsilon_m$.  In the presence of a periodic potential
with the same lattice periodicity, the energy levels will
disperse with ${\bf k}$.   This leads to a band structure
that looks identical to that of an ordinary insulator.

\subsubsection{The TKNN invariant}
\label{sec:tknn}

What is the difference between a
quantum Hall state characterized by \eqref{qhall} and an ordinary insulator?
The answer, explained in a seminal \citeyear{thouless82} paper by
Thouless, Kohmoto, Nightingale and den Nijs(TKNN) is a matter of topology.
A 2D band structure consists of a
mapping from the crystal momentum ${\bf k}$ (defined on a torus)
to the Bloch Hamiltonian ${\cal H}({\bf k})$.
Gapped band structures can
be classified topologically by considering the equivalence classes of
${\cal H}({\bf k})$ that can be continuously deformed into one another
without closing the energy gap.  These classes are distinguished by a
topological invariant $n \in \mathbb{Z}$ ($\mathbb{Z}$ denotes the integers)
called the Chern invariant.

The Chern invariant is rooted in the mathematical
theory of fiber bundles\cite{nakahara90}, but it
can be understood physically in terms of the Berry phase\cite{berry84} associated
with Bloch wavefunctions $|u_m({\bf k})\rangle$.  Provided
there are no accidental degeneracies, when ${\bf k}$ is transported around a
closed loop, $|u_m({\bf k})\rangle$ acquires a well defined Berry phase given by the
line integral of ${\cal A}_m = i\langle u_m|\nabla_k|u_m\rangle$.
This may be expressed as a surface integral of the Berry
flux, ${\cal F}_m = \nabla\times{\cal A}_m$.  The Chern invariant is the total
Berry flux in the Brillouin zone,
\begin{equation}
n_m = \frac{1}{2\pi} \int d^2{\bf k} {\cal F}_m.
 \label{chern}
\end{equation}
$n_m$ is integer quantized for reasons analogous to the quantization of the
Dirac magnetic monopole.  The total Chern number, summed over all occupied bands,
$n = \sum_{m=1}^N n_m$
is invariant even if there are degeneracies between
occupied bands, provided the gap separating occupied and empty bands remains
finite.
TKNN showed that $\sigma_{xy}$, computed using the Kubo formula
has the same form, so that $N$ in \eqref{qhall} is identical to
$n$.  The Chern number $n$ is a topological
invariant in the sense that it can not change when the Hamiltonian
varies smoothly.   This helps
to explain the robust quantization of $\sigma_{xy}$.


The meaning of \eqref{chern} can be clarified by a simple analogy.  Rather than
maps from the Brillioun zone to a Hilbert
space, consider simpler maps from 2D to 3D, which describe surfaces.  2D surfaces can be
topologically classified by their genus, $g$, which counts the number of
holes.  For instance, a sphere (Fig. \ref{fig:insulator}(c)) has $g=0$,
while a donut (Fig. \ref{fig:insulator}(f)) has $g=1$.
A beautiful theorem in mathematics due to Gauss and Bonnet \cite{nakahara90} states
that the integral of the Gaussian
curvature over a closed surface is a quantized topological invariant, and
its value is related to $g$.  The Chern number is an integral
of a related curvature.

\subsubsection{Graphene, Dirac electrons, Haldane model}
\label{sec:dirac}

A simple example of the quantum Hall effect in a band theory
is provided by a  model of graphene in a
periodic magnetic field introduced by \textcite{haldane88}.  We will briefly digress
here to introduce graphene because it will provide insight into the conception of the
2D quantum spin Hall insulator, and because the physics of Dirac electrons
present in graphene has important parallels at the surface of a 3D topological insulator.

Graphene is a 2D form of carbon that is of high current interest
\cite{novoselov05, zhangy05, geim07,neto09}.  What makes graphene interesting
electronically is the fact that the conduction band and valence
band touch each other at two distinct points in the Brillouin zone.  Near
those points the electronic dispersion resembles the linear dispersion
of massless relativistic particles, described by the Dirac equation
\cite{divencenzo84,semenoff84}.
The simplest description of graphene employs a two band model for
the $p_z$ orbitals on the two equivalent atoms in the unit
cell of graphene's honeycomb lattice.  The Bloch Hamiltonian is
then a $2\times 2$ matrix,
\begin{equation}
{\cal H}({\bf k}) = {\bf h}({\bf k}) \cdot \vec \sigma,
\label{dirac}
\end{equation}
where $\vec\sigma = (\sigma_x,\sigma_y,\sigma_z)$ are Pauli matrices
and ${\bf h}({\bf k}) = (h_x({\bf k}),h_y({\bf k}),0)$.  The
combination of inversion ($\cal P$) and time reversal ($\cal T$) symmetry requires
$h_z({\bf k})=0$ because
$\cal P$ takes $h_z({\bf k})$ to $-h_z(-{\bf k})$, while $\cal T$ takes
$h_z({\bf k})$ to $+h_z(-{\bf k})$.  The Dirac points
occur because the two component ${\bf h}({\bf k})$ can have point
zeros in two dimensions.  In graphene they occur at two points,
${\bf K}$ and ${\bf K}'=-{\bf K}$, whose locations at the
Brillouin zone corners are fixed
by graphene's rotational symmetry.  For small ${\bf q} \equiv {\bf k}-{\bf K}$,
${\bf h}({\bf q}) = \hbar v_F {\bf q}$, where $v_F$ is a velocity,
so ${\cal H}({\bf q}) = \hbar v_F {\bf q}\cdot\vec\sigma$
has the form of a 2D massless Dirac Hamiltonian.

The degeneracy at the Dirac point is protected by $\cal P$ and $\cal T$ symmetry.
By breaking these symmetries the degeneracy
can be lifted.  For instance, $\cal P$ symmetry is violated if the
two atoms in the unit cell are inequivalent.  This allows
$h_z({\bf k})$ to be non zero.  If $h_z({\bf k})$ is small, then
near ${\bf K}$ \eqref{dirac} becomes a massive Dirac
Hamiltonian,
\begin{equation}
{\cal H}({\bf q}) = \hbar v_F{\bf q}\cdot\vec\sigma + m \sigma_z
\label{diracmass}
\end{equation}
where $m = h_z({\bf K})$.  The dispersion
$E({\bf q}) = \pm \sqrt{|\hbar v_F {\bf q}|^2+m^2}$ has an
energy gap $2|m|$ .  Note
that ${\cal T}$ symmetry requires the Dirac point at ${\bf K}'$ has
a mass $m'=h_z({\bf K}')$ with the same magnitude {\it and} sign, $m' = m$.  This state
describes an ordinary insulator.

\textcite{haldane88} imagined lifting the degeneracy by breaking ${\cal T}$
symmetry with a magnetic field that is zero
on the average, but has the full symmetry the lattice.  This
perturbation allows nonzero $h_z({\bf k})$ and introduces a mass to
the Dirac points.  However, $\cal P$ symmetry requires the masses at ${\bf
K}$ and ${\bf K}'$
have {\it opposite} sign, $m' = - m$. Haldane showed that this gapped
state is not an insulator, but rather a quantum Hall state with
$\sigma_{xy} = e^2/h$.

This non-zero Hall conductivity can be understood in terms of
\eqref{chern}.  For a two level Hamiltonian of the form of \eqref{dirac} it
is well known that the Berry flux\cite{berry84} is related to
the solid angle subtended by the unit vector $\hat h({\bf k}) =
{\bf h}({\bf k})/|{\bf h}({\bf k})|$, so that \eqref{chern} takes the form
\begin{equation}
n = \frac{1}{4\pi}\int d^2{\bf k}
(\partial_{k_x} \hat h \times \partial_{k_y} \hat h)\cdot \hat h .
\end{equation}
This simply counts the number of times $\hat h({\bf k})$ wraps around
the unit sphere as a function of ${\bf k}$.  When the masses $m=m'=0$
$\hat h({\bf k})$ is confined to the equator $h_z=0$, with a unit
(and opposite)
winding around each of the Dirac points where $|{\bf h}|=0$.
For small but finite $m$, $|{\bf h}|\ne 0$ everywhere, and $\hat h({\bf K})$
visits the north or south pole, depending on the sign of $m$.
It follows that each Dirac point contributes $\pm e^2/2h$ to $\sigma_{xy}$.  In
the insulating state with $m=m'$ the two cancel, so
$\sigma_{xy}=0$. In the quantum Hall state they add.

It is essential that there were an {\it even} number of Dirac points, since
otherwise the Hall conductivity would be quantized to a half integer.  This is
in fact guaranteed by the {\it fermion doubling theorem} \cite{nielssen83}, which states that for
a ${\cal T}$ invariant system Dirac points must come in pairs.  We will
return to this issue in section \ref{sec:3d}, where the surface of a topological
insulator provides a loophole for this theorem.

\subsubsection{Edge states and the bulk-boundary correspondence}
\label{sec:edge}

\begin{figure}
\includegraphics[width=3in]{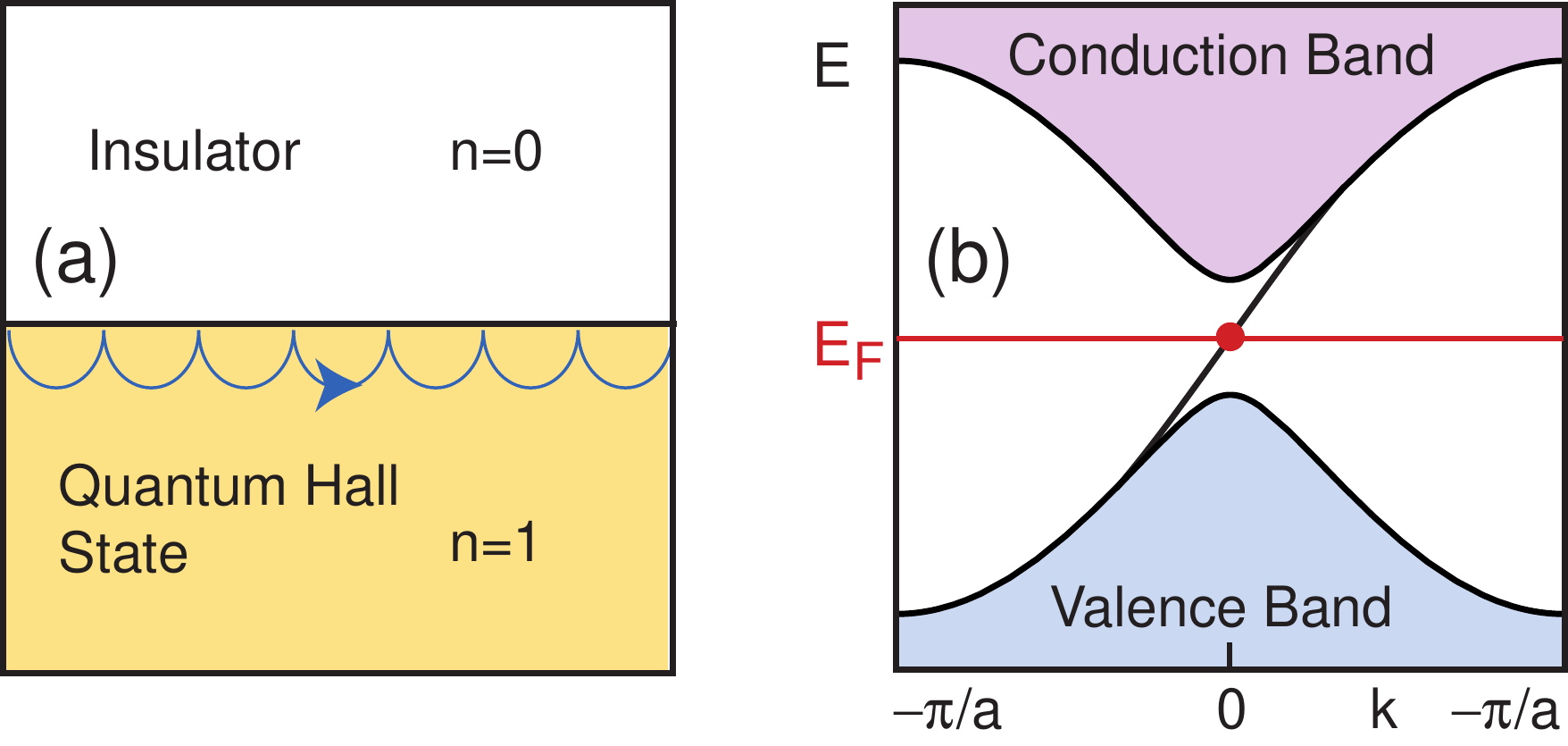}
\caption{The interface between a quantum Hall state and an insulator has chiral edge mode.
(a) depicts the skipping cyclotron orbits.  (b) shows the electronic structure of a semi infinite
strip described by the Haldane model.  A single edge state connects the valence band to the
conduction band.}
\label{fig:qhalledge}
\end{figure}

A fundamental consequence of the topological classification of gapped
band structures is the existence of gapless conducting states at
interfaces where the topological invariant changes.  Such edge states
are well known at the interface between the integer quantum Hall
state and vacuum \cite{halperin82}.  They may be understood in terms of the
skipping motion electrons execute as their
cyclotron orbits bounce off the edge (Fig. \ref{fig:qhalledge}(a)).
Importantly, the electronic
states responsible for this motion are {\it chiral} in the sense that
they propagate in one direction only along the edge.  These states
are insensitive to disorder because there are no states available for
backscattering -- a fact that underlies the perfectly
quantized electronic transport in the quantum Hall effect.

The existence
of such ``one way" edge states is deeply related to the topology of
the bulk quantum Hall state.
Imagine an interface where a crystal slowly interpolates as a
function of distance $y$
between a quantum Hall state ($n=1$) and a trivial insulator
($n=0$).  Somewhere along the way the energy gap has to vanish,
because otherwise it is impossible for the topological invariant to
change.  There will therefore be low energy electronic states bound
to the region where the energy gap passes through zero.  This interplay
between topology and gapless modes is
ubiquitous in physics, and has appeared in many contexts.
It was originally found by \textcite{jackiw76} in their analysis of
a 1D field theory.  Similar ideas were used by
\textcite{su79} to describe soliton states in polyacetalene.

A simple theory of the chiral edge states based on \textcite{jackiw76}
can be developed using the
two band Dirac model \eqref{diracmass}.  Consider an interface where
the mass $m$ at one of the Dirac points changes sign as a function of $y$.
We thus let $m \rightarrow m(y)$, where $m(y)>0$ gives the
insulator for $y>0$ and $m(y)<0$ gives the quantum Hall state for $y<0$.
Assume $m'>0$ is fixed.  The
Schr\"odinger equation, obtained by replacing ${\bf q}$ by
$-i \vec\nabla$ in \eqref{diracmass}, has a simple and elegant exact
solution,
\begin{equation}
\psi_{q_x}(x,y) \propto e^{i q_x x} e^{-\int_0^y dy' m(y') dy'/v_F}
\left(\begin{array}{c} 1\\1 \end{array}\right),
\label{jackiwrebbi}
\end{equation}
with $E(q_x) = \hbar v_F q_x$.  This band of states intersects the
Fermi energy $E_F$ with a positive group velocity $dE/dq_x = \hbar
v_F$ and defines a right moving chiral edge mode.

In the 1980's related ideas were applied to narrow gap
semiconductors,
which can be modeled using a 3D massive Dirac Hamiltonian\cite{volkov85,fradkin86}.
An interface where the Dirac mass changes sign is associated with gapless 2D
Dirac fermion states.  These share some similarities with
the surface states of a 3D topological insulator, but as we shall see in
section \ref{sec:strongweak}, there is a fundamental difference.
In a separate development, \textcite{kaplan92} showed that in lattice
quantum chromodynamics 4D chiral fermions could be simulated on
a 5D lattice by introducing a similar domain wall.  This provided a method for circumventing the
doubling theorem\cite{nielssen83}, which prevented the simulation of
chiral fermions on a 4D lattice.  Quantum Hall edge states and surface states
of a topological insulator evade similar doubling theorems.

The chiral edge states in the quantum Hall effect can be seen explicitly by solving
the Haldane model in a semi-infinite geometry with an edge at $y=0$.  Fig.
\ref{fig:qhalledge}(b) shows the energy levels as a function of the momentum
$k_x$ along the edge.  The solid regions show the bulk conduction and
valence bands, which form continuum states and show the energy gap
near ${\bf K}$ and ${\bf K}'$.  A single band, describing states
bound to the edge connects the
valence band to the conduction band with a positive group velocity.

By changing the Hamiltonian near the surface the dispersion
of the edge states can be modified.  For instance, $E(q_x)$ could
develop a kink so that the edge states intersect $E_F$
three times -- twice with a positive group velocity and once with a
negative group velocity.  The difference $N_R-N_L$ between the number of right
and left moving modes, however,
can not change, and is determined by the topological structure of
the bulk states.  This is summarized by the {\it bulk-boundary
correspondence}:
\begin{equation}
N_R-N_L = \Delta n,
\label{bulkboundary}
\end{equation}
where $\Delta n$ is the difference in the Chern number across the interface.

\subsection{$Z_2$ topological insulator}
\label{sec:z2topo}

Since the Hall conductivity
is odd under ${\cal T}$, the topologically non trivial
states described in the preceding section
can only occur when ${\cal T}$ symmetry is broken.
However, the spin orbit interaction allows
a {\it different} topological class of insulating band structures when ${\cal T}$
 symmetry is unbroken \cite{kanemele05a}.  The key to understanding this new
topological class is to examine the role of ${\cal T}$ symmetry
for spin 1/2 particles.

${\cal T}$ symmetry is represented by an antiunitary operator
$\Theta = \exp(i\pi S_y/\hbar) K$, where $S_y$ is the spin
operator and $K$ is complex conjugation.
For spin 1/2 electrons, $\Theta$ has the property $\Theta^2 = -1$.
This leads to an important constraint, known as Kramers' theorem,
that all eigenstates of a ${\cal T}$ invariant Hamiltonian
are at least twofold degenerate.
This follows because if a non degenerate state $|\chi\rangle$ existed then
$\Theta |\chi\rangle = c |\chi\rangle$ for some constant $c$.  This would mean $\Theta^2
|\chi\rangle = |c|^2 |\chi\rangle$, which is not allowed because
$|c|^2 \ne -1$.  In the absence of spin orbit interactions, Kramers'
degeneracy is simply the degeneracy between up and down spins.  In
the presence of spin orbit interactions, however, it has nontrivial
consequences.

A ${\cal T}$ invariant Bloch Hamiltonian must satisfy
\begin{equation}
\Theta {\cal H}({\bf k}) \Theta^{-1} = {\cal H}(-{\bf k}).
\label{trsym}
\end{equation}
One can classify the equivalence classes of
Hamiltonians satisfying this constraint that can be smoothly deformed
without closing the energy gap.  The TKNN invariant is $n=0$, but
there is an additional invariant with
two possible values $\nu = 0$ or $1$ \cite{kanemele05b}.   The fact that there are two
topological classes can be understood
by appealing to the bulk-boundary correspondence.

\begin{figure}
\includegraphics[width=3in]{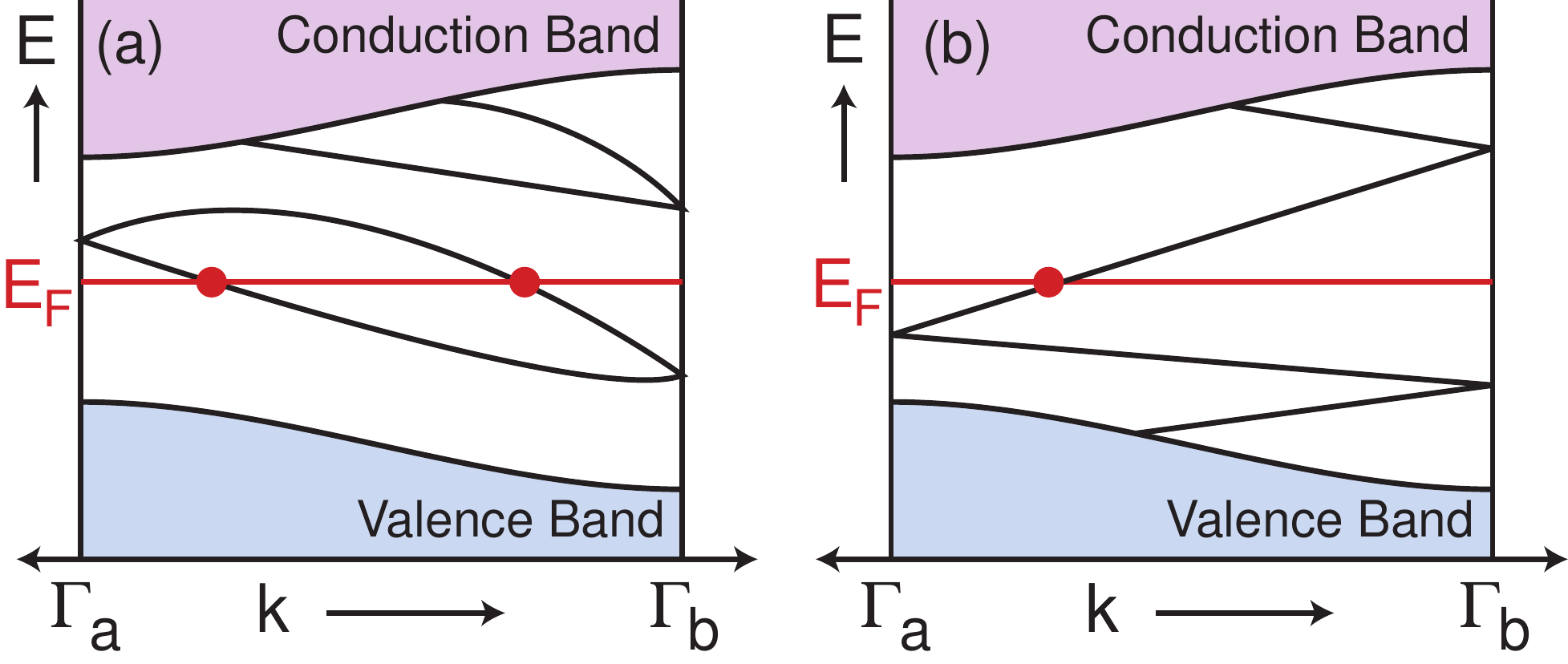}
\caption{Electronic dispersion between two boundary Kramers degenerate points
$\Gamma_a=0$ and $\Gamma_b=\pi/a$.  In (a) the number of surface states crossing the Fermi
energy $E_F$ is even, whereas in (b) it is odd.
An odd number of crossings leads to topologically protected metallic boundary states.}
\label{fig:zigzag}
\end{figure}

In Fig. \ref{fig:zigzag} we show plots analogous to Fig. \ref{fig:qhalledge} showing the electronic
states associated with the edge of a ${\cal T}$ invariant 2D
insulator as a function of the crystal momentum along the edge.  Only
half of the Brillouin zone $0<k_x<\pi/a$ is shown because ${\cal T}$
symmetry requires that the other
half $-\pi/a<k<0$ is a mirror image.  As in Fig. \ref{fig:qhalledge}, the
shaded regions depict the bulk conduction and valence bands separated by an
energy gap.  Depending on the details of the Hamiltonian near the
edge there may or may not be states bound to the edge inside the gap.
If they are present, however, then Kramers
theorem requires they be twofold degenerate at the ${\cal T}$
invariant momenta $k_x=0$ and $k_x=\pi/a$ (which is the same as
$-\pi/a$). Away from these special points, labeled $\Gamma_{a,b}$ in
Fig. \ref{fig:zigzag}, a spin orbit
interaction will split the degeneracy.  There are two
ways the states at $k_x=0$ and $k_x=\pi/a$ can connect.  In
Fig \ref{fig:zigzag}(a) they connect pairwise.  In this case
the edge states can be eliminated by pushing all of the bound
states out of the gap. Between $k_x=0$ and $k_x=\pi/a$, the bands intersect
$E_F$ an even number of times. In contrast, in Fig. \ref{fig:zigzag}b the
edge states cannot be eliminated.  The bands intersect $E_F$
an odd number of times.

Which of these alternatives occurs depends on
the topological class of the bulk band structure.  Since each band intersecting
$E_F$ at $k_x$ has a Kramers partner at $-k_x$, the bulk-boundary
correspondence relates the number $N_K$ of Kramers pairs of edge
modes intersecting $E_F$ to the change in the $\mathbb{Z}_2$
invariants across the interface,
\begin{equation}
N_K = \Delta \nu \ {\rm mod} \ 2.
\end{equation}
We conclude that a 2D topological insulator has
topologically protected edge states.  These form a unique
1D conductor, whose properties will be discussed in section
\ref{sec:qshi}.  The above considerations can be generalized to
3D topological insulators, discussed in section \ref{sec:3d},
which have protected surface states.

There are several mathematical formulations of the $\mathbb{Z}_2$ invariant $\nu$
\cite{kanemele05b,fukane06,moorebalents07,fukane07,fukui07,fukui08,qihugheszhang08,roy09a,wang09}.  One
approach \cite{fukane06} is to define a unitary matrix
$w_{mn}({\bf k}) = \langle u_m({\bf k})|\Theta|u_n(-{\bf k})\rangle$
built from the occupied Bloch functions
$|u_m({\bf k})\rangle$.  Since $\Theta$ is anti unitary
and $\Theta^2=-1$, $w^T({\bf k}) = - w(-{\bf k})$.  There are four
special points $\Lambda_a$ in the bulk 2D Brillouin zone where ${\bf k}$ and
$-{\bf k}$ coincide, so $w(\Lambda_a)$ is
antisymmetric.  The determinant of an antisymmetric matrix is the square of
its pfaffian, which allows us to define
$\delta_a = {\rm Pf}[w(\Lambda_a)]/\sqrt{{\rm Det}[w(\Lambda_a)]} = \pm 1$.
{\it Provided} $|u_m({\bf k})\rangle$ is chosen continuously
throughout the Brillouin zone (which is always possible), the branch
of the square root can be specified globally, and the $\mathbb{Z}_2$
invariant is
\begin{equation}
(-1)^\nu = \prod_{a=1}^4 \delta_a.
\label{deltaa}
\end{equation}
This formulation can be
generalized to 3D topological insulators, and involves
the 8 special points in the 3D Brillouin zone.

The calculation of $\nu$ is simpler
if the crystal has extra symmetry.  For instance, if the 2D system
conserves the perpendicular spin $S_z$, then the
up and down spins have independent
Chern integers $n_\uparrow$, $n_\downarrow$.
${\cal T}$ symmetry requires $n_\uparrow+n_\downarrow = 0$, but
the difference $n_\sigma = (n_\uparrow-n_\downarrow)/2$ defines a quantized spin
Hall conductivity \cite{sheng06}.  The $\mathbb{Z}_2$ invariant is then simply
\begin{equation}
\nu = n_\sigma \ {\rm mod} \ 2.
\end{equation}
While $n_\uparrow$, $n_\downarrow$ lose
their meaning when $S_z$ non conserving terms (which
are inevitably present) are added, $\nu$ retains its identity.

If the crystal has inversion symmetry there is another shortcut to
computing $\nu$ \cite{fukane07}.  At the special points $\Lambda_a$
the Bloch states $u_m(\Lambda_a)$ are also parity eigenstates with eigenvalue
$\xi_m(\Lambda_a)=\pm 1$.  The $\mathbb{Z}_2$ invariant then simply
follows from \eqref{deltaa} with
\begin{equation}
\delta_a = \prod_{m} \xi_m(\Lambda_a),
\label{parityz2}
\end{equation}
where the product is over the Kramers pairs of occupied bands.
This has proven useful for identifying
topological insulators from band structure calculations
\cite{fukane07,teofukane08,zhangh09,pesin10,guo09}.

\subsection{Topological superconductor, Majorana fermions}
\label{sec:superconductor}

Considerations of topological band theory can also be used to
topologically classify superconductors.  This is a subject that has seen
fascinating recent theoretical
developments \cite{schnyder08,kitaev09,qihughesraduzhang09,roy08}.
We will give an introduction that focuses on the simplest
model superconductors.  The more
general case will be briefly touched on at the end.  This section will provide
the conceptual basis for topological superconductors and explain
the emergence of Majorana fermions in superconducting
systems.  It will also provide background for section \ref{sec:proximity}, where we
discuss Majorana states in superconductor-topological
insulator structures along with possible applications to topological
quantum computing.  Readers who wish to skip the discussion of superconductivity
can proceed directly to section \ref{sec:qshi}.

\subsubsection{Bogoliubov de Gennes theory}
\label{sec:bdg}

In the BCS mean field theory of a superconductor the Hamiltonian for a system
of spinless electrons may be written in the form \cite{degennes66},
\begin{equation}
H-\mu N =
\frac{1}{2}\sum_{\bf k} \left(\begin{array}{cc} c_{\bf k}^\dagger & c_{-{\bf k}} \end{array}\right)
{\cal H}_{BdG}({\bf k})
\left(\begin{array}{c} c_{\bf k} \\ c_{-{\bf k}}^\dagger \end{array}\right)
\label{bdg}
\end{equation}
where $c_{\bf k}^\dagger$ is an electron creation operator and
${\cal H}_{BdG}$ is a $2 \times 2$ block matrix, which in Nambu's notation
may be written in terms of Pauli matrices $\vec\tau$ as
\begin{equation}
{\cal H}_{BdG}({\bf k}) = ({\cal H}_0({\bf k})-\mu) \tau_z + \Delta_1({\bf k})
\tau_x + \Delta_2({\bf k}) \tau_y.
\end{equation}
Here ${\cal H}_0({\bf k})$ is the Bloch Hamiltonian in the absence of
superconductivity and $\Delta = \Delta_1 + i\Delta_2$ is the BCS mean
field pairing potential, which for spinless particles must have odd parity,
$\Delta(-{\bf k}) = -\Delta({\bf k})$.
For a uniform system the excitation spectrum
of a superconductor is given by the eigenvalues of ${\cal
H}_{BdG}$, which exhibit a superconducting energy gap.  More generally, for
spatially dependent ${\cal H}_0$ and
$\Delta$ the Schr\"odinger equation associated with ${\cal H}_{BdG}$ is
known as the Bogoliubov de Gennes (BdG) equation.

Since \eqref{bdg} has both $c$ and $c^\dagger$ on both sides there is
an inherent redundancy built into the BdG Hamiltonian.  For
$\Delta=0$, ${\cal H}_{BdG}$ includes
two copies of ${\cal H}_0$ with opposite sign.  More generally, ${\cal H}_{BdG}$
has an intrinsic {\it particle-hole symmetry} expressed by
\begin{equation}
\Xi {\cal H}_{BdG}({\bf k}) \Xi^{-1} = - {\cal H}_{BdG}(-{\bf k}),
\label{phsym}
\end{equation}
where the particle-hole operator, $\Xi = \tau_x K$,
satisfies $\Xi^2 = +1$.  \eqref{phsym} follows from ${\cal H}_0(-{\bf k})={\cal H}_0({\bf k})^*$
and the odd parity of the real $\Delta({\bf k})$.  It follows that every
eigenstate of ${\cal H}_{BdG}$ with energy $E$ has a partner at
$-E$.  These two states are redundant because the
Bogoliubov quasiparticle operators associated with them satisfy
$\Gamma_E^\dagger = \Gamma_{-E}$.  Thus, creating a quasiparticle in state
$E$ has the same effect as removing one from state $-E$.

The particle-hole symmetry constraint \eqref{phsym} has a similar
structure to the time reversal constraint in \eqref{trsym}, so it is
natural to consider the classes of BdG Hamiltonians that
can be continuously deformed into one another without closing the
energy gap.  In the simplest case, spinless fermions, the classification
can be shown to be $\mathbb{Z}_2$
in one dimension and $\mathbb{Z}$ in two dimensions.
As in section \ref{sec:z2topo}, this can be most easily understood by appealing to
the bulk-boundary correspondence.

\subsubsection{Majorana fermion boundary states}
\label{sec:majoranaedge}

\begin{figure}
\includegraphics[width=3in]{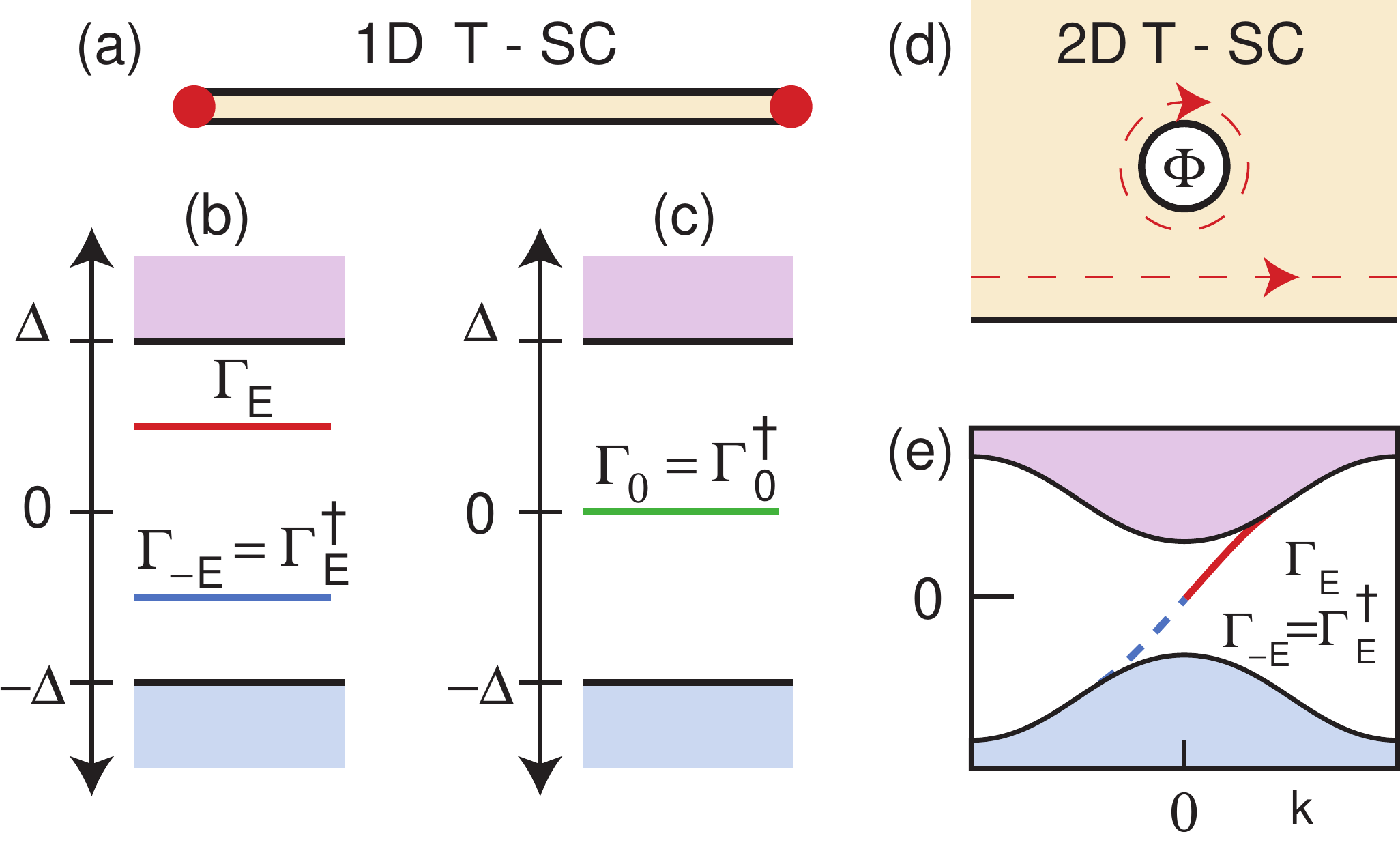}
\caption{Boundary states for a topological superconductor (T-SC).  (a) shows a 1D superconductor
with bound states at its ends.  (b,c) show the end state spectrum for an ordinary 1D superconductor
(b) and a 1D topological superconductor (c).  (d) shows a topological 2D superconductor
with a chiral Majorana edge mode (e).  A vortex with flux $\Phi = h/2e$ is associated
with a zero mode (c).}
\label{fig:scedge}
\end{figure}

At the end of a 1D superconductor \cite{kitaev00} there may or may not
be discrete states within the energy gap that are bound to the end
(Fig. \ref{fig:scedge}(a-c)).
If they are present, then every state at $+E$ has a partner at
$-E$.  Such finite energy pairs are not topologically protected
because they can simply be pushed out of the energy gap.  However,
a single {\it unpaired} bound state at $E=0$ {\it is} protected because it can't move away
from $E=0$.  The presence or absence of such a zero mode is determined by the
$\mathbb{Z}_2$ topological class of the bulk 1D superconductor.

The Bogoliubov quasiparticle states associated with the zero modes are
fascinating objects \cite{nayak08,readgreen00,stern04,ivanov01,kitaev00}.  Due to the particle-hole redundancy
the quasiparticle operators satisfy $\Gamma_0 = \Gamma_0^\dagger$.
Thus, a quasiparticle is its own antiparticle -- the defining
feature of a {\it Majorana fermion}.
A Majorana fermion is essentially {\it half} of an ordinary
Dirac fermion.  Due to the particle-hole redundancy, a single fermionic state
is associated with each {\it pair} of $\pm E$ energy levels.  The presence or absence
of a fermion in this state defines a two level system with energy splitting $E$.
Majorana zero modes
must always come in pairs (for instance, a 1D superconductor has two ends), and a
well separated pair
defines a {\it degenerate} two level system, whose quantum state is stored nonlocally.
This has profound implications, which we will return to in section
\ref{sec:proximity}, when we discuss the proposal by \textcite{kitaev03} to use these
properties for quantum information processing.

In two dimensions the integer classification, $\mathbb{Z}$,
gives the number of {\it chiral Majorana} edge
modes (Fig. \ref{fig:scedge}(d,e)), which resemble chiral modes in the quantum Hall effect, but for
the particle-hole redundancy.  A spinless superconductor with $p_x+ip_y$ symmetry
is the simplest model 2D topological superconductor.
Such superconductors will also exhibit Majorana bound states at the core of vortices
\cite{caroli64,volovik99,readgreen00}.
This may be understood simply by considering the vortex to be a hole
in the superconductor circled by an edge mode (Fig. \ref{fig:scedge}(d)).
When the flux in the
hole is $h/2e$ the edge modes are quantized such that one state is exactly
at $E=0$.

Majorana fermions have been studied in particle
physics for decades, but have not been definitively observed \cite{majorana37,wilczek09}.
A neutrino {\it might} be a Majorana fermion.  Efforts to observe
certain lepton number violating
neutrinoless double $\beta$ decay processes may resolve that issue \cite{avignone08}.
In condensed matter physics, Majorana fermions can arise due to
a paired condensate that allows a pair of fermionic quasiparticles to
``disappear" into the condensate.  They have been predicted in a
number of physical systems related to the spinless
$p_x+i p_y$ superconductor, including the Moore-Read state of the
$\nu=5/2$ quantum Hall effect \cite{mooreread91,greiter92,readgreen00}, Sr$_2$RuO$_4$ \cite{dassarma06},
cold fermionic atoms near a Feshbach resonance \cite{gurarie05,tewari07} and
2D structures that combine superconductivity, magnetism and strong
spin orbit coupling \cite{sato09,sau10,lee09}.
In Section Vb we will discuss the prospect for creating
Majorana fermion states at interfaces between topological
insulators and ordinary superconductors \cite{fukane08}.

\subsubsection{Periodic table}
\label{sec:periodic}

Topological insulators and superconductors fit together into a rich and elegant
mathematical structure that generalizes the notions of topological band
theory described above \cite{schnyder08,schnyder09,kitaev09,ryu09}.  The classes of equivalent Hamiltonians are determined by specifying the
symmetry class and the dimensionality.  The symmetry class depends on
the presence or absence of ${\cal T}$ symmetry \eqref{trsym} with $\Theta^2=\pm 1$ and/or
particle-hole symmetry \eqref{phsym} with $\Xi^2=\pm 1$.  There are 10 distinct classes,
which are closely related to the \textcite{altland97} classification of random
matrices.  The topological classifications, given by
$\mathbb{Z}$, $\mathbb{Z}_2$ or $0$, show a regular pattern as a function of
symmetry class and dimensionality and can be arranged into the {\it periodic
table} of topological insulators and superconductors shown in Table \ref{tab:periodic}.

\begin{table}
  \centering
\begin{ruledtabular}
\begin{tabular}{c|ccc|cccccccc}
\multicolumn{4}{c|}{Symmetry } & \multicolumn{8}{c}{ $d$} \\
\multicolumn{1}{c}{AZ} &$\hspace{1.5mm}\Theta\hspace{1.5mm} $ &
                        $\hspace{1.5mm} \Xi\hspace{1.5mm} $ &
                        $\hspace{1.5mm} \Pi\hspace{1.5mm} $ &
 $1$   &  $2$ &  $3$ &  $4$ &  $5$ & $6$ & $7$& $8$ \\
 \hline
A & $0$ & $0$ & $0$  &$0$& $\mathbb{Z}$ &$0$& $\mathbb{Z}$ &$0$& $\mathbb{Z}$ &$0$& $\mathbb{Z}$\\
AIII & $0$ & $0$ & $1$ & $\mathbb{Z}$ &$0$& $\mathbb{Z}$ &$0$& $\mathbb{Z}$ &$0$& $\mathbb{Z}$& $0$\\
\hline
 AI & $1$ & $0$ & $0$  &$0$&$0$&$0$&$\mathbb{Z}$&$0$&$\mathbb{Z}_2$&$\mathbb{Z}_2$& $\mathbb{Z}$ \\
 BDI & $1$ &$1$ &$1$ & $\mathbb{Z}$ &$0$&$0$&$0$&$\mathbb{Z}$&$0$&$\mathbb{Z}_2$& $\mathbb{Z}_2$\\
 D & $0$ &$1$ &$0$ & $\mathbb{Z}_2$& $\mathbb{Z}$ &$0$&$0$&$0$&$\mathbb{Z}$&$0$&$\mathbb{Z}_2$\\
 DIII&$-1$ &$1$ &$1$ &$\mathbb{Z}_2$& $\mathbb{Z}_2$& $\mathbb{Z}$ &$0$&$0$&$0$&$\mathbb{Z}$&$0$\\
 AII & $-1$ & $0$ & $0$ &$0$&$\mathbb{Z}_2$& $\mathbb{Z}_2$& $\mathbb{Z}$ &$0$&$0$& $0$&$\mathbb{Z}$\\
 CII & $-1$ &$-1$ & $1$&$\mathbb{Z}$ & $0$&$\mathbb{Z}_2$& $\mathbb{Z}_2$& $\mathbb{Z}$ &$0$&$0$&$0$ \\
 C & $0$ & $-1$& $0$ & $0$ &$\mathbb{Z}$ &$0$&$\mathbb{Z}_2$& $\mathbb{Z}_2$& $\mathbb{Z}$ &$0$& $0$\\
 CI & $1$ & $-1$ & $1$& $0$ & $0$&$\mathbb{Z}$&$0$&$\mathbb{Z}_2$& $\mathbb{Z}_2$& $\mathbb{Z}$& $0$ \\

\end{tabular}
\end{ruledtabular}
\caption{Periodic table of topological insulators and superconductors.  The 10 symmetry
classes are labeled using the notation of \textcite{altland97} (AZ) and are
specified by presence or absence of ${\cal T}$ symmetry
$\Theta$, particle-hole symmetry $\Xi$ and chiral symmetry $\Pi = \Xi\Theta$.
$\pm 1$ and $0$ denotes the presence and absence of symmetry,
with $\pm 1$ specifying the value of $\Theta^2$ and
$\Xi^2$.  As a function of symmetry and space dimensionality, $d$, the topological
classifications ($\mathbb{Z}$, $\mathbb{Z}_2$ and $0$) show a regular pattern
that repeats when $d \rightarrow d+8$.  }
\label{tab:periodic}
\end{table}

The quantum Hall state (Class A, no symmetry; $d=2$), the $\mathbb{Z}_2$ topological
insulators (Class AII, $\Theta^2=-1$; $d=2,3$) and the $\mathbb{Z}_2$ and $\mathbb{Z}$ topological
superconductors (Class D, $\Xi^2=1$; $d=1,2$) described above are each entries in the periodic table.
There are also other non trivial entries describing different topological
superconducting and superfluid phases.  Each non trivial phase is predicted,
via the bulk-boundary correspondence to have gapless boundary states.  One
notable example is superfluid $^3$He B \cite{schnyder08,roy08,nagato09,qihughesraduzhang09,volovik03,volovik09}, in
(Class DIII, $\Theta^2=-1$, $\Xi^2=+1$; $d=3$) which has a $\mathbb{Z}$
classification, along with gapless 2D Majorana fermion modes on its
surface.   A generalization of the quantum Hall state introduced by
\textcite{zhang01} corresponds to the $d=4$ entry in class A or AII.
There are also other entries
in physical dimensions that have yet to
be filled by realistic systems.  The search is on to discover such
phases.

\section{Quantum Spin Hall Insulator}
\label{sec:qshi}

The 2D topological insulator is known as a quantum spin
Hall insulator.  This state was originally theorized to exist in
graphene \cite{kanemele05a} and in 2D semiconductor systems with a uniform
strain gradient \cite{bernevig06}.  It was subsequently predicted to exist \cite{bernevighugheszhang06}, and was
then observed \cite{konig07}, in HgCdTe quantum well structures.  In section \ref{sec:graphene} we
will introduce the physics of this state in the model graphene system
and describe its novel edge states.  Section \ref{sec:hgcdte} will
review the experiments, which have also been the subject of the review article
by \textcite{konig08}.

\subsection{Model system: graphene}
\label{sec:graphene}

\begin{figure}
\includegraphics[width=3in]{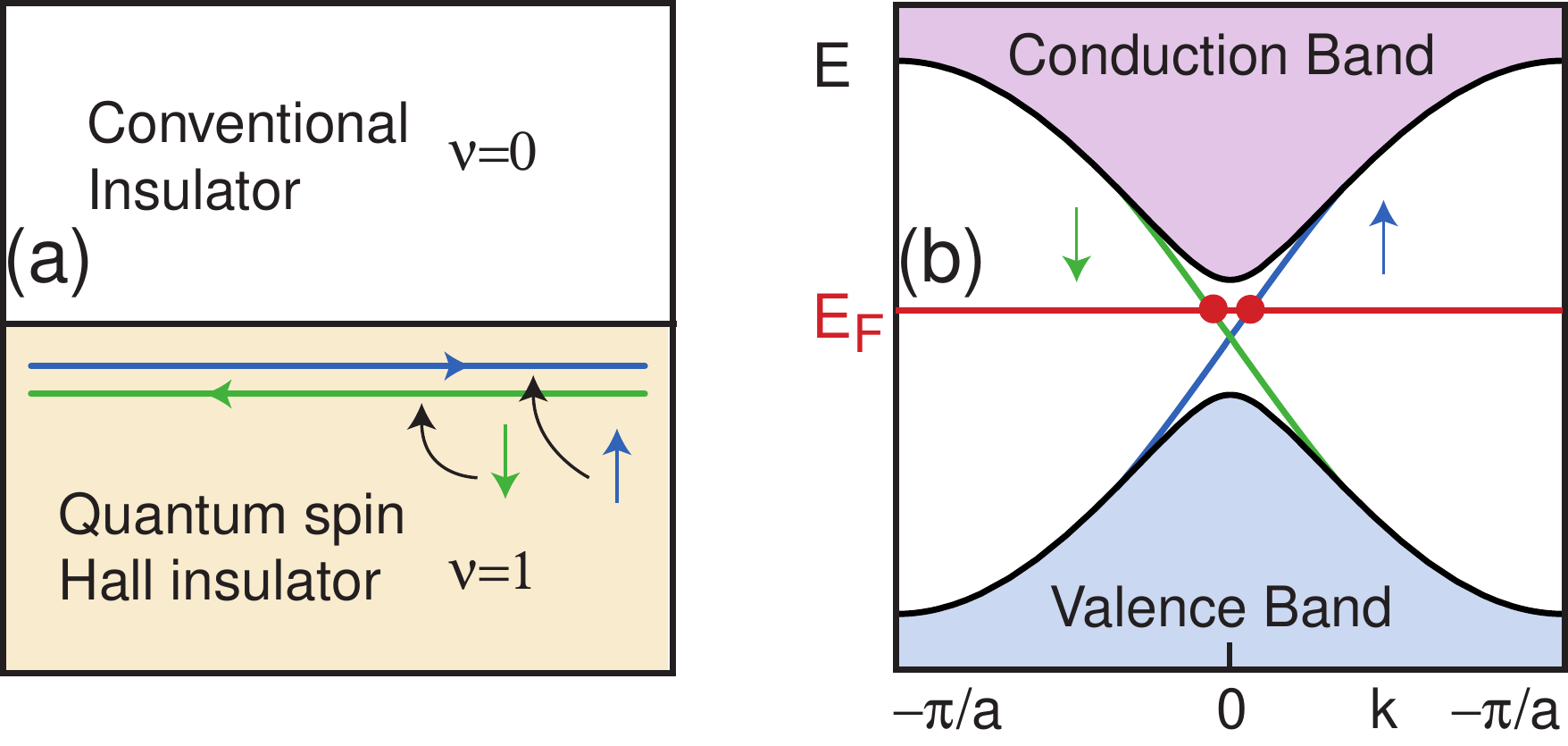}
\caption{Edge states in the quantum spin Hall insulator.  (a) shows the interface between a
QSHI and an ordinary insulator, and (b) shows the edge state dispersion in the graphene model,
in which up and down spins propagate in opposite directions.}
\label{fig:qshedge}
\end{figure}

In section \ref{sec:dirac} we argued that the degeneracy at the Dirac
point in graphene is protected by inversion and ${\cal T}$
symmetry.  That argument ignored the spin of the
electrons.  The spin orbit interaction allows a new mass
term in \eqref{dirac} that respects {\it all} of graphene's
symmetries.
In the simplest picture, the intrinsic spin orbit interaction
commutes with the electron spin $S_z$, so the Hamiltonian decouples
into two independent Hamiltonians for the up and down spins.  The
resulting theory is simply two copies the \textcite{haldane88} model
with opposite signs of the Hall conductivity for up and down spins.
This does not violate ${\cal T}$ symmetry because time reversal
flips both the spin and $\sigma_{xy}$.  In an applied
electric field, the up and down spins have Hall currents that flow in
opposite directions.  The Hall conductivity is thus zero, but there
is a quantized {\it spin Hall} conductivity, defined by
$J_x^\uparrow-J_x^\downarrow = \sigma_{xy}^s E_y$ with
$\sigma_{xy}^s = e/2\pi$ --
a quantum spin Hall effect.  Related ideas were mentioned in
earlier work on the planar state of $^3$He films\cite{volovik89}.
Since it is two copies a quantum Hall state, the quantum
spin Hall state must have gapless edge states (Fig. \ref{fig:qshedge}).

The above discussion was predicated on the conservation of
spin, $S_z$.  This is not a fundamental symmetry, though, and spin
non conserving processes -- present in any real system -- invalidate the
meaning of $\sigma_{xy}^s$.  This brings into
question theories that relied on spin conservation to predict an
integer quantized $\sigma_{xy}^s$ \cite{volovik89,bernevig06,qi06}, as well as the
influential theory of the (non quantized) spin Hall insulator \cite{murakami04}.
\textcite{kanemele05a} showed that due to ${\cal T}$ symmetry
the edge states in the quantum spin Hall insulator are robust
even when spin conservation is violated because their crossing at $k=0$ is
protected by the Kramers degeneracy discussed in section
\ref{sec:z2topo}.  This established the quantum spin Hall insulator as a
topological phase.

The quantum spin Hall edge states have the
important ``spin filtered" property that the up spins propagate in one direction,
while the down spins propagate in the other.  Such edge states were later dubbed
``helical" \cite{wu06}, in analogy with the correlation between spin and momentum
of a particle known as helicity.
They form a unique 1D conductor that is essentially {\it half} of an ordinary 1D conductor.
Ordinary conductors, which have both up and down spins propagating in
both directions, are fragile because the electronic states are
susceptible to Anderson localization in the presence of weak
disorder \cite{anderson58,lee85}.  By contrast, the quantum spin Hall edge states can not be localized,
even for strong disorder.  To see this, imagine an edge that is
disordered in a finite region, and perfectly clean outside that
region.  The exact eigenstates can be determined by solving the
scattering problem relating incoming waves to those reflected from
and transmitted through the disordered region.  \textcite{kanemele05a}
showed that the reflection amplitude is {\it odd} under ${\cal T}$
-- roughly because it involves flipping the spin.  It
follows that unless ${\cal T}$ symmetry is broken, an incident
electron is transmitted {\it perfectly} across the disordered region.
Thus, eigenstates at any energy are extended, and
at temperature $T=0$, the edge state transport is
ballistic.  For $T>0$ inelastic backscattering processes are
allowed, which will in general lead to a finite conductivity.

The edge states are similarly protected from the
effects of weak electron interactions, though for strong interactions Luttinger liquid
effects lead to a magnetic instability \cite{wu06,xu06}.   This
strongly interacting phase is interesting because it will
exhibit charge $e/2$ quasiparticles similar to
solitons in the \textcite{su79} model.  For sufficiently strong interactions
similar fractionalization could
be observed by measuring shot noise in the presence of magnetic
impurities \cite{maciejko09} or at a quantum point contact \cite{teokane09}.

\subsection{HgTe/CdTe quantum well structures}
\label{sec:hgcdte}

Graphene is made out of carbon -- a
light element with a weak spin orbit interaction.  Though there
is disagreement on its absolute magnitude \cite{min06,huertas06,yao07,boettger07,gmitra09},
the energy gap in graphene is likely to be small.
Clearly, a better place to look for this physics would be in
materials with strong spin orbit interactions, made from
heavy elements near the bottom of the periodic table.
To this end, \textcite{bernevighugheszhang06} (BHZ) had the brilliant idea to
consider quantum well structures of HgCdTe.  This paved
the way to the experimental discovery of the quantum spin Hall
insulator phase.


Hg$_{1-x}$Cd$_x$Te is a family of semiconductors with strong spin orbit
interactions \cite{dornhaus83}.  CdTe has a band structure similar to other
semiconductors.  The conduction band edge states have an $s$ like symmetry,
while the valence band edge states have a $p$ like symmetry.
In HgTe, the $p$ levels rise above the $s$ levels, leading to an {\it inverted}
band structure.  BHZ considered a quantum well structure where HgTe is
sandwiched between layers of CdTe.  When the thickness of the HgTe layer is
$d < d_c= 6.3$ nm the 2D electronic states bound to the quantum well have the
normal band order.  For $d > d_c$, however, the 2D bands
invert.  BHZ showed that the inversion of the bands as a function of increasing $d$ signals a
quantum phase transition between the trivial insulator and the quantum spin
Hall insulator.   This can be understood simply in the approximation that
the system has inversion symmetry.  In this case, since the $s$ states
and $p$ states have opposite parity the bands will cross each other at $d_c$ without
an avoided crossing.  Thus the energy gap at $d=d_c$ vanishes.
From \eqref{parityz2}, the change in the parity of
the valence band-edge state signals a phase transition in which
the $\mathbb{Z}_2$ invariant $\nu$ changes.

\begin{figure}
\includegraphics[width=3in]{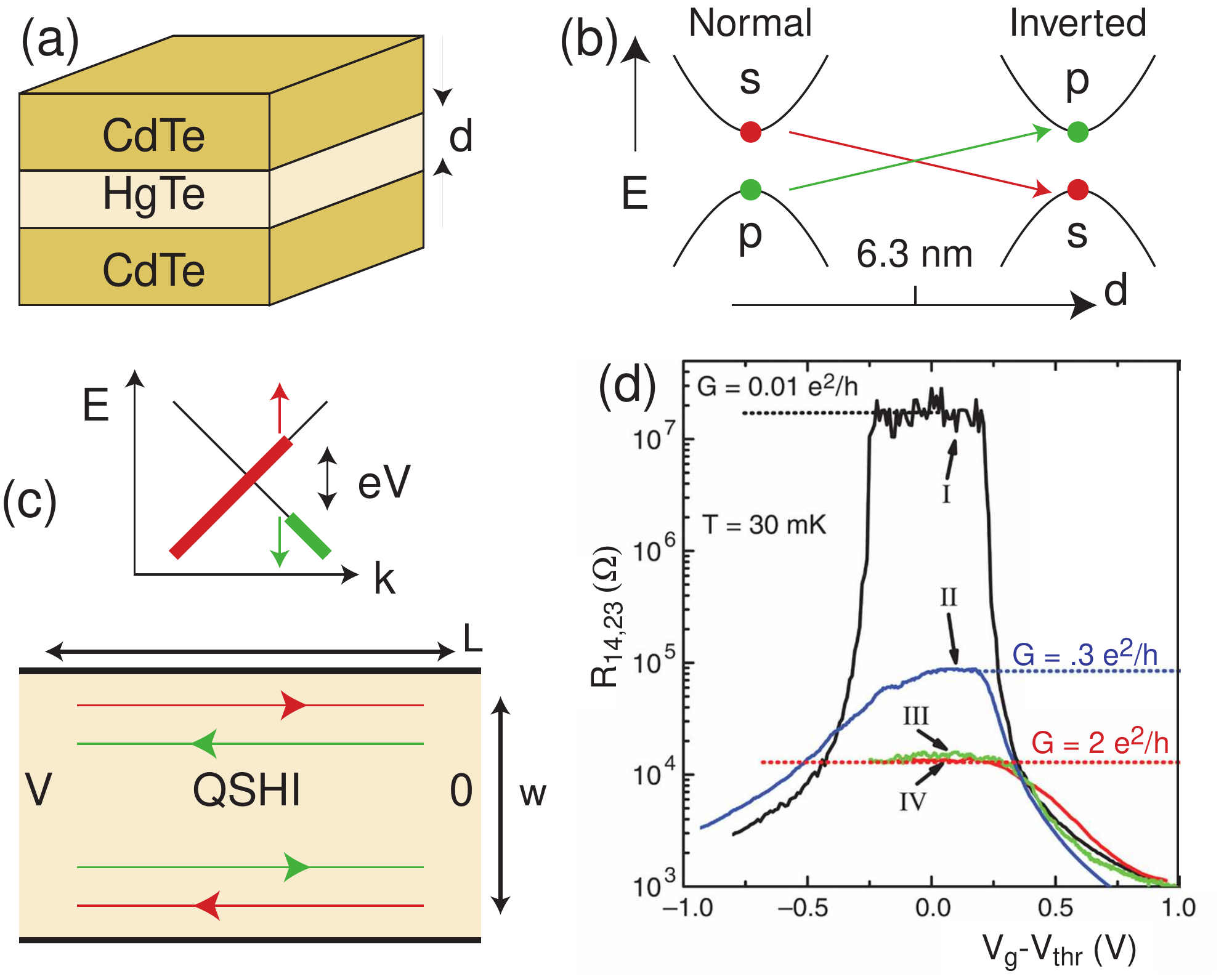}
\caption{(a) A HgCdTe quantum well structure.  (b) As a function of layer thickness
$d$ the 2D quantum well states cross at a band inversion transition.  The
inverted state is the QSHI, which has helical edge states (c) that
have a non equilibrium population determined by the leads.  (d) shows
experimental two terminal conductance as a function of a gate voltage that
tunes $E_F$ through the bulk gap.    Sample I, with $d<d_c$ shows insulating
behavior, while samples III and IV show quantized transport associated with
edge states.  Adapted from \onlinecite{konig07}.  Reprinted with permission from AAAS.}
\label{fig:hgcdtefig}
\end{figure}


Within a year of the theoretical proposal the W\"urzburg group, led by Laurens
Molenkamp, made the devices and performed transport experiments that
showed the first signature of the quantum spin Hall insulator.
\textcite{konig07}
measured the electrical conductance due to the edge states.
The low temperature ballistic edge state transport
can be understood within a simple Landauer-B\"uttiker
\cite{buttiker88}
framework in which the edge states are populated according to the
chemical potential of the lead that they emanate from.  This leads to
a quantized conductance $e^2/h$ associated with each set of edge
states.  Fig. \ref{fig:hgcdtefig}(d) shows the resistance measurements for a series of
samples as a function of a gate voltage which tunes the Fermi energy
through the bulk energy gap.  Sample I is a narrow quantum well
that has a large resistance in the gap.  Samples II, III and IV are
wider wells in the inverted regime.   Samples III and IV exhibit a
conductance $2e^2/h$ associated with the top and bottom edges.  Samples III and IV have
the same length $L = 1\mu$ but different widths
$w = 0.5\mu,1\mu$, indicating transport is at the edge.  Sample
II  ($L=20\mu$) showed finite temperature scattering effects. These
experiments convincingly demonstrate the existence of the edge
states of the quantum spin Hall insulator.  Subsequent
experiments have established the inherently nonlocal electronic
transport in the edge states \cite{roth09}.

\section{3D Topological Insulators}
\label{sec:3d}

In the summer of 2006 three groups of theorists independently discovered that the
topological characterization of the quantum spin Hall insulator state
has a natural generalization in three dimensions \cite{fukanemele07,moorebalents07,roy09b}.
\textcite{moorebalents07} coined the term ``topological
insulator" to describe this electronic phase.  \textcite{fukanemele07} established the
connection between the bulk topological order and the presence of unique conducting surface
states.  Soon after, this phase was predicted in several real materials \cite{fukane07}, including
Bi$_{1-x}$Sb$_x$ as well as strained HgTe and $\alpha-$Sn.  In 2008, \textcite{hsieh08}
reported the experimental discovery of the first 3D topological
insulator in Bi$_{1-x}$Sb$_x$.  In 2009 ``second generation" topological
insulators, including Bi$_2$Se$_3$, which has numerous desirable properties,
were identified experimentally \cite{xia09a} and theoretically \cite{xia09a,zhangh09}.
In this section we will review these developments.

\subsection{Strong and weak topological insulators}
\label{sec:strongweak}

\begin{figure}
\includegraphics[width=3in]{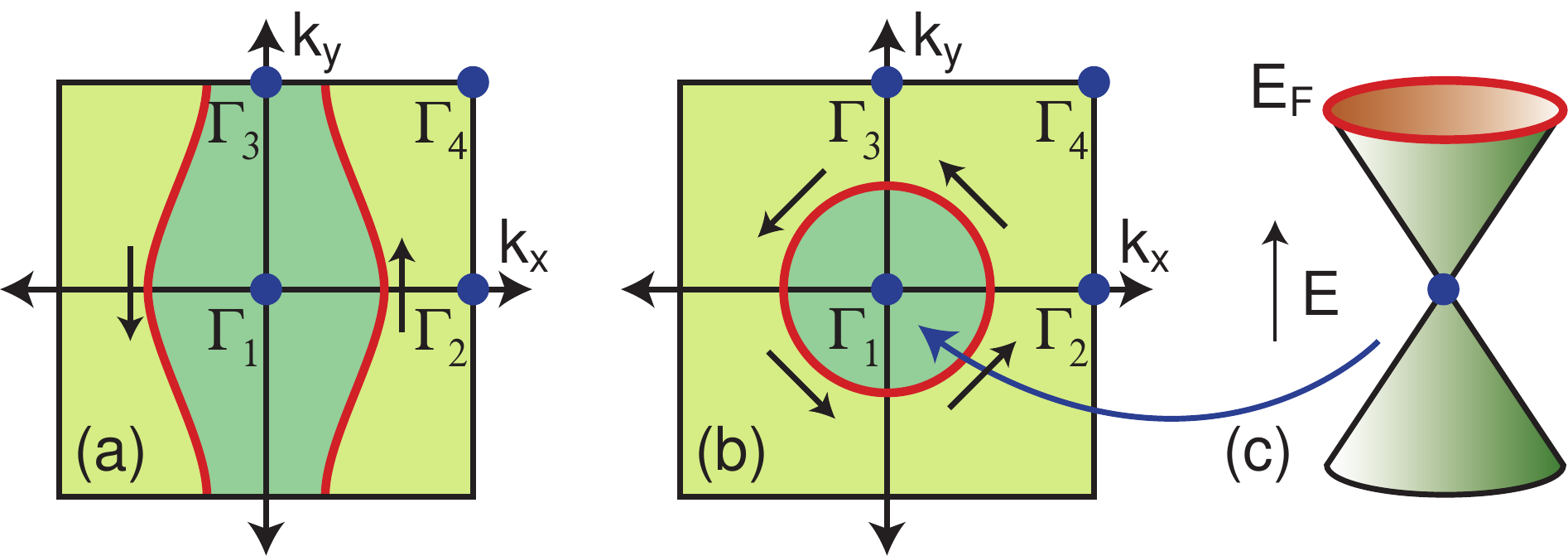}
\caption{Fermi circles in the surface Brillouin zone for (a) a weak topological insulator and
(b) a strong topological insulator.  In the simplest strong topological insulator the Fermi circle
encloses a single Dirac point (c).}
\label{fig:surfacebz}
\end{figure}

A 3D topological insulator is characterized by four $\mathbb{Z}_2$
topological invariants $(\nu_0;\nu_1\nu_2\nu_3)$ \cite{fukanemele07,moorebalents07,roy09b}.
They can be most
easily understood by appealing to the bulk-boundary correspondence,
discussed in section \ref{sec:z2topo}.
The surface states of a 3D crystal
can be labeled with a 2D crystal
momentum.  There are four ${\cal T}$ invariant points $\Gamma_{1,2,3,4}$ in the
surface Brillouin zone, where surface states, if present, must be
Kramers degenerate (Fig. \ref{fig:surfacebz}(a,b)).  Away from these special points, the spin orbit
interaction will lift the degeneracy.  These Kramers degenerate
points therefore form 2D {\it Dirac points} in the surface band
structure (Fig. \ref{fig:surfacebz}(c)).
The interesting question is how the Dirac points at the
different ${\cal T}$ invariant points connect to each other.
Between any pair $\Gamma_a$ and $\Gamma_b$, the surface state structure will resemble
either Fig. \ref{fig:zigzag}a or \ref{fig:zigzag}b.  This determines whether the surface Fermi
surface intersects a line joining $\Gamma_a$ to $\Gamma_b$ an even
or an odd number of times.  If it is odd, then the surface states are
topologically protected.  Which of these two alternatives occurs is
determined by the four bulk $\mathbb{Z}_2$ invariants.

The simplest non trivial 3D topological insulators may be
constructed by stacking layers of the 2D quantum spin Hall
insulator.  This is analogous to a similar construction for 3D
integer quantum Hall states \cite{kohmoto92}.
The helical edge states of the layers
then become anisotropic surface states.
A possible surface Fermi surface for weakly coupled layers stacked along the
$y$ direction is sketched in Fig. \ref{fig:surfacebz}(a).  In this figure a single
surface band intersects the Fermi energy between $\Gamma_1$ and $\Gamma_2$ and
between $\Gamma_3$ and $\Gamma_4$, leading to the non trivial connectivity in
Fig. \ref{fig:zigzag}(b).  This layered state is referred to as a weak topological
insulator, and has $\nu_0=0$.
The indices $(\nu_1\nu_2\nu_3)$ can be interpreted as Miller indices
describing the orientation of the layers.
Unlike the 2D helical edge states of a single layer, ${\cal T}$
symmetry does not protect these surface states.  Though the
surface states must be present for a clean surface, they
can be localized in the presence of disorder.
Interestingly, however, a line {\it dislocation} in
a weak topological insulator is associated with protected 1D
helical edge states \cite{ran09}.

$\nu_0=1$ identifies a distinct phase, called a strong topological insulator,
which can not be interpreted as a
descendent of the 2D quantum spin Hall insulator.  $\nu_0$ determines
whether an even or an odd number
of Kramers points is enclosed by the surface Fermi circle.
In a strong topological insulator the surface Fermi circle encloses an {\it odd} number of
Kramers degenerate Dirac points.    The simplest case, with a single Dirac
point(Fig. \ref{fig:surfacebz}(b,c)), can be described by the Hamiltonian,
\begin{equation}
{\cal H}_{\rm surface} = -i\hbar v_F \vec\sigma\cdot\vec\nabla,
\label{surfacedirac}
\end{equation}
where $\vec\sigma$ characterizes the spin.  (For a surface with a mirror plane,
symmetry requires $\vec S \propto \hat z\times\vec\sigma$.)

The surface electronic structure of a topological insulator is
similar to graphene, except rather than having four
Dirac points (2 valley $\times$ 2 spin) there is just a single Dirac
point.  This appears to violate the fermion doubling theorem \cite{nielssen83}
discussed in section \ref{sec:dirac}.  The resolution is that the partner
Dirac points reside on {\it opposite} surfaces.

The surface states of a strong topological insulator form a unique
2D topological metal \cite{fukanemele07,fukane07} that is essentially {\it half}
an ordinary metal.  Unlike an ordinary metal,
which has up and down spins at every point on the Fermi surface, the surface
states are not spin degenerate.  Since ${\cal T}$ symmetry requires that states at
momenta ${\bf k}$ and $-{\bf k}$ have opposite spin, the spin must rotate with
${\bf k}$ around the Fermi surface, as indicated in Fig.
\ref{fig:surfacebz}(b).  This leads to a non trivial Berry phase
acquired by an electron going around the Fermi
circle.  ${\cal T}$ symmetry requires that this phase be $0$ or
$\pi$.  When an electron circles a Dirac point, its spin rotates by
$2\pi$, which leads to a $\pi$ Berry phase.

The Berry phase has
important consequences for the behavior in a magnetic field (to be
discussed in section \ref{sec:qhetopomag}) and for the effects of disorder.  In
particular, in an ordinary 2D electron gas the electrical
conductivity decreases with decreasing temperature, reflecting the
tendency towards Anderson localization in the presence of disorder \cite{lee85}.
The $\pi$ Berry
phase changes the sign of the weak localization correction to the
conductivity leading to weak {\it antilocalization} \cite{suzuura02}.  In fact, the
electrons at the surface of a strong topological insulator can not be
localized even for strong disorder, as long as the bulk energy gap
remains intact \cite{nomura07}.  In this regard, the situation is similar to the edge states of the
quantum spin Hall insulator discussed in section \ref{sec:graphene}, however,
the electron motion on the surface is diffusive rather than ballistic.

The Dirac surface states \eqref{surfacedirac} can be
understood in a 3D Dirac theory\cite{qihugheszhang08} where the Dirac mass changes sign at the
surface, analogous to \eqref{jackiwrebbi}.  Such domain wall states were first
discussed for Pb$_{1-x}$Sn$_x$Te\cite{volkov85,fradkin86},
which exhibits a band inversion as a function of
$x$.  An appropriate interface where $x$ changes was
predicted to have 2D gapless states.  There is an important difference
between these interface states and the surface
states of a topological insulator, though, because the band inversion in
Pb$_{1-x}$Sn$_x$Te occurs at 4 equivalent valleys.  Since 4 is even, PbTe and SnTe are
both trivial insulators.  The interface states are not
topologically protected from disorder in the sense discussed above.
However, if the valleys can be split by applying uniaxial stress, then
the topological insulator can occur in the vicinity of the band inversion
transition\cite{fukane07}.
Related ideas were also applied to interfaces between HgTe and CdTe
\cite{chang85,cade85,linliu85,pankratovpakhomov87}.
In this case, the band inversion occurs in a single valley, but since
HgTe is a zero gap semiconductor, the surface states are not
protected.  Nonetheless, if the cubic symmetry of the bulk HgTe can
be lifted by applying uniaxial stress, a gap can be introduced in HgTe,
so the HgTe-CdTe interface will have topologically protected states\cite{fukane07}.

\subsection{The first 3D topological insulator: Bi$_{1-x}$Sb$_x$}
\label{sec:bisb}

\begin{figure}
\includegraphics[width=3.2in]{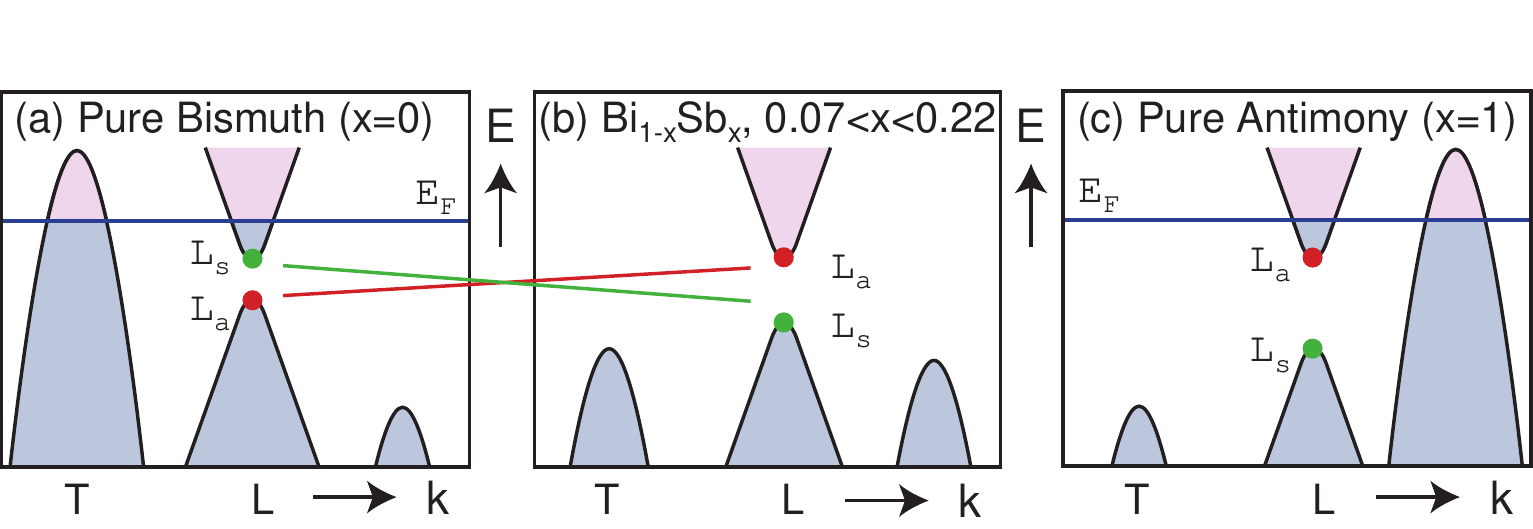}
\caption{Schematic representation of the band structure of Bi$_{1-x}$Sb$_x$,
which evolves from semimetallic behavior for $x<.07$ to semiconducting behavior
for $.07<x<.22$ and back to semimetallic behavior for $x>.18$.  The conduction
and valence bands $L_{s,a}$ invert at $x \sim .04$. }
\label{fig:bisbbands}
\end{figure}
\begin{table}
  \centering
\begin{ruledtabular}
\begin{tabular}{c|ccccc|c||c|ccccc|c}
\multicolumn{7}{c||}{Bi: Class $(0;000)$ } &
\multicolumn{7}{c}{Sb: Class $(1;111)$ } \\
\multicolumn{1}{c}{$\Lambda_a$} &
\multicolumn{5}{c}{Symmetry label}&
\multicolumn{1}{c||}{$\delta_a$} &
\multicolumn{1}{c}{$\Lambda_a$} &
\multicolumn{5}{c}{Symmetry label} &
\multicolumn{1}{c}{$\delta_a$} \\
\hline
$1\Gamma$ & $\Gamma_6^+$ & $\Gamma_6^-$ &$\Gamma_6^+$ &$\Gamma_6^+$ &
$\Gamma_{45}^+$ & $-1$ &
$1\Gamma$ & $\Gamma_6^+$ & $\Gamma_6^-$ &$\Gamma_6^+$ &$\Gamma_6^+$ &
$\Gamma_{45}^+$ & $-1$ \\
$3L$ & $L_s$ &$L_a$ &$L_s$ &$L_a$ &$L_a$ & $-1$ &
$3L$ & $L_s$ &$L_a$ &$L_s$ &$L_a$ &$L_s$ & $+1$ \\
$3X$ & $X_a$ & $X_s$ & $X_s$ & $X_a$ & $X_a$  & $-1$ &
$3X$ & $X_a$ & $X_s$ & $X_s$ & $X_a$ & $X_a$  & $-1$ \\
$1T$ & $T_6^-$ & $T_6^+$ & $T_6^-$ & $T_6^+$ & $T_{45}^-$ & $-1$ &
$1T$ & $T_6^-$ & $T_6^+$ & $T_6^-$ & $T_6^+$ & $T_{45}^-$ & $-1$\\
\end{tabular}
\end{ruledtabular}
\caption{Symmetry labels for the Bloch states at the 8 ${\cal T}$ invariant momenta
$\Lambda_a$ for the 5 valence bands of Bi and Sb.
$\delta_a$ are given by \eqref{parityz2} and
determine the topological class $(\nu_0;\nu_1\nu_2\nu_3)$ by
relations similar to \eqref{deltaa}.
The difference between Bi and Sb is due to the inversion of
the $L_s$ and $L_a$ bands that occurs at $x \sim .04$.}
\label{tab:bisbtab}
\end{table}

The first 3D topological insulator to be identified experimentally
was the semiconducting alloy
Bi$_{1-x}$Sb$_{x}$, whose unusual surface bands were mapped in an
angle resolved photoemission spectroscopy (ARPES) experiment by a Princeton
University group led by Hasan\cite{hsieh08}.

Bismuth antimony alloys have long been studied for their
thermoelectric properties \cite{lenoir96}.  Pure bismuth is a semimetal with
strong spin-orbit interactions.  Its band structure, depicted schematically in
Fig. \ref{fig:bisbbands}(a) features conduction and valence bands that overlap,
leading to pockets of holes near the $T$ point in the Brillouin zone
and pockets of electrons near the three equivalent $L$ points.
The valence and conduction bands at the $L$ point, derived
from antisymmetric ($L_a$) and symmetric ($L_s$) orbitals have a small
energy gap $\Delta$.  The states near $L$ have a nearly linear
dispersion that is well described by a $3+1$
dimensional Dirac equation \cite{wolff64} with a small mass.  These facts have been used to
explain many peculiar properties of bismuth.

Substituting bismuth with antimony changes the critical energies of
the band structure (Fig. \ref{fig:bisbbands}(b)). At an Sb concentration of $x
\approx .04$, the gap $\Delta$ between $L_a$ and $L_s$ closes and a
truly massless 3D Dirac point is realized. As $x$ is
further increased this gap reopens with an inverted ordering.
For $x >.07$ the top of the valence band at $T$ moves below the bottom of the conduction
band at $L$, and the material becomes an
insulator. Once the band at $T$ drops below the valence band at $L$, at
$x\sim .09$, the system is a direct gap insulator with
a massive Dirac like bulk bands.  As $x$ is increased further, the conduction
and valence bands remain separated, and for $x \gtrsim .22$ the valence band at
a different point rises above the conduction band, restoring the semimetallic
state.

Since pure bismuth and pure antimony both have a finite direct band gap, their
valence bands can be topologically classified.  Moreover, since they have
inversion symmetry, Eq. \ref{parityz2} can be used to determine the topological indices.
Table \ref{tab:bisbtab} shows the symmetry labels that specify the parity
of the Bloch states, for the occupied bands at the 8
${\cal T}$ invariant points in the bulk Brillouin zone \cite{liuallen95}.
\textcite{fukane07} used this information to deduce
that bismuth is in the trivial $(0;000)$ class, while antimony is in the
$(1;111)$ class.  Since the semiconducting alloy is on the antimony side of the
band inversion transition, it is predicted to inherit the $(1;111)$ class from
antimony.

Charge transport experiments,
which were successful for identifying the 2D topological
insulator \cite{konig07}, are problematic in 3D materials because
the signature in the conductivity
of the topological character of the surface states is more subtle in 3D.
Moreover, it is difficult to separate the surface contribution to the conductivity
from that of the bulk.
Angle resolved photoemission spectroscopy (ARPES) is an ideal tool for probing
the topological character of the surface states.
ARPES uses a photon to eject an electron from a
crystal, then determines the surface or bulk electronic structure
from an analysis of the momentum of the emitted electron.
High-resolution ARPES performed with
modulated photon energy allows for a clear isolation of surface
states from that of the bulk 3D band-structure because surface states
do not disperse along a direction perpendicular to the surface where
as the bulk states do. Moreover, unlike in a transport experiment,
ARPES carried out in a spin resolution mode can, in addition, measure
the distribution of spin orientations on the Fermi surface which can
be used to estimate the Berry phase on the surface. Spin
sensitivity is critically important for probing the existence of
spin-momentum locking on the surface expected as a consequence of
bulk topological order.

\begin{figure}
\includegraphics[width=2.8in]{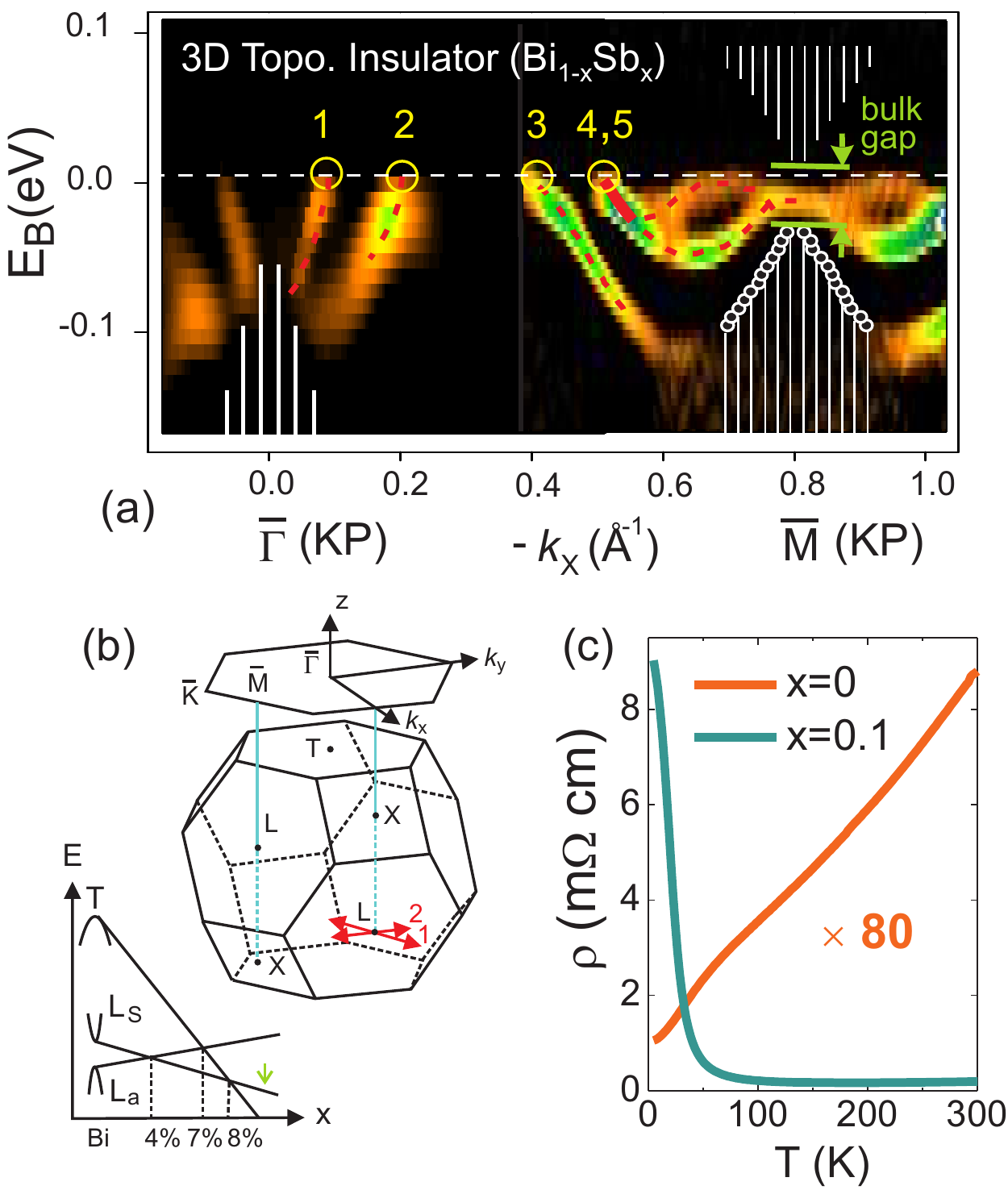}
\caption{Topological surface states in
Bi$_{1-x}$Sb$_{x}$:  (a) ARPES data on the 111 surface of
Bi$_{0.9}$Sb$_{0.1}$ which probes
the occupied surface states as a function of momentum on the line connecting
the ${\cal T}$ invariant points $\bar\Gamma$ and $\bar M$ in the surface Brillouin
zone.  Only the surface bands cross the Fermi energy 5 times.  This, along with
further detailed ARPES results \cite{hsieh08} establish that the semiconducting
alloy Bi$_{1-x}$Sb$_{x}$ is a strong topological insulator in the $(1;111)$
class.  (b) shows a schematic of the 3D Brillouin zone and its (111) surface
projection.  (c) contrasts the resistivity of semimetallic pure Bi with
the semiconducting alloy.  Adapted from \onlinecite{hsieh08}.}
\label{fig:zfig1} \end{figure}

Experiments by \textcite{hsieh08} probed both the bulk and
surface electronic structure of Bi$_{.09}$Sb$_{.91}$ with ARPES.
Fig. \ref{fig:zfig1}(a) shows the ARPES spectrum, which can be interpreted
as a map of the energy of the occupied electronic states as a function
of momentum along the line connecting $\bar\Gamma$ to $\bar M$ in the projected
surface Brillouin zone (Fig. \ref{fig:zfig1}(b)).
Bulk energy bands associated with the $L$ point are observed that reflect the
nearly linear 3D Dirac like dispersion.
The same experiments observed several surface states that span the bulk gap.

The observed surface state structure of Bi$_{1-x}$Sb$_x$ has similarities with the
surface states in pure Bi, which have been studied
previously \cite{patthey94,agergaard01,ast01,hirahara06,hofmann06}.  In pure Bi, two
bands emerge from the bulk band continuum near $\bar\Gamma$ to form a central
electron pocket and an adjacent hole lobe.  These two bands result from the
spin splitting of a surface state, and are thus expected to be singly
degenerate.  In Bi$_{1-x}$Sb$_x$, there are additional states near $\bar M$,
which play a crucial role.

As explained in Section \ref{sec:strongweak}, Kramers' theorem requires surface states to be
doubly degenerate at the ${\cal T}$ invariant points $\bar\Gamma$ and each of the
three equivalent $\bar M$ points.
Such a Kramers point is indeed observed at $\bar M$ approximately $15\pm 5$meV below
$E_F$.  As expected for a system
with strong spin orbit interactions, the degeneracy is lifted away from $\bar M$.
The observed surface bands cross the Fermi energy 5 times between $\bar\Gamma$ and
$\bar M$.  This odd number of crossings is analogous to Fig. \ref{fig:zigzag}(b),
and indicates that these surface states are topologically protected.
Accounting for the threefold rotational symmetry and mirror symmetry of the
111 surface, this data shows that the surface Fermi surface encloses
$\bar\Gamma$ an odd number of times, while it encloses the three equivalent
$\bar M$ points an even number of times.  This establishes Bi$_{1-x}$Sb$_x$ as
a strong topological
insulator, with $\nu_0 = 1$.  The data is consistent with the predicted
$(1;111)$ topological class.

\begin{figure}
\includegraphics[width=3.3in]{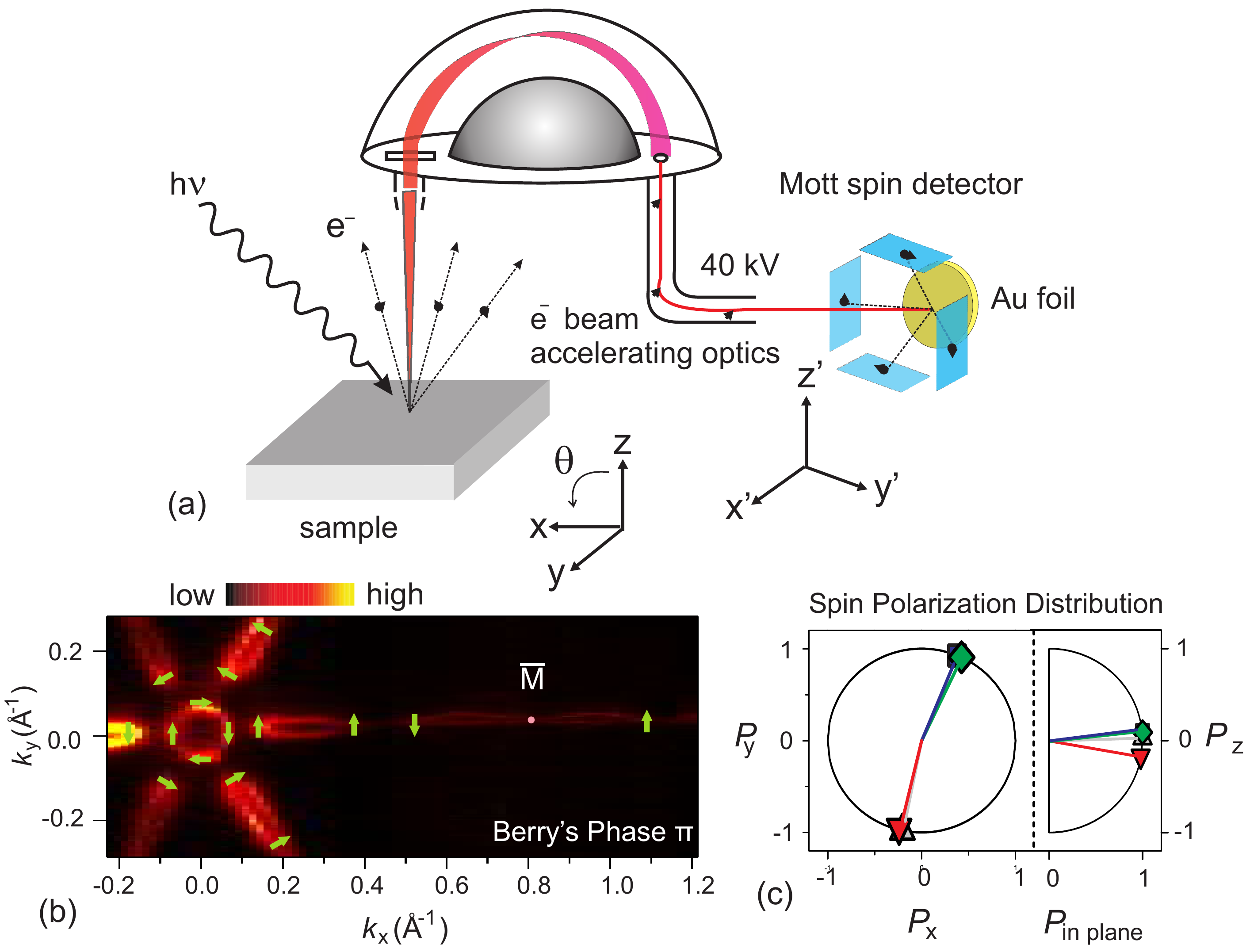}
\caption{Topological spin-textures:
Spin resolved photoemission directly probes the non trivial spin textures
of the topological insulator surface.
(a) A schematic of spin-ARPES
measurement set up that was used to measure the spin
distribution on the (111) surface Fermi surface of
Bi$_{0.91}$Sb$_{0.09}$.  (b) Spin orientations on the surface create a
vortex like pattern around $\Gamma$-point. A net Berry phase
$\pi$ is extracted from the full Fermi surface data.
(c) Net polarization along x-, y- and z- directions are shown.
\textit{P$_z$}$\sim$0 suggests that spins lie mostly within the
surface plane.  Adapted from \onlinecite{hsieh09a,hsieh09d,hsieh10}.  }
\label{fig:zfig2} \end{figure}

A distinguishing feature of topological insulator surface states is the
intimate correlation between spin and momentum they exhibit, which underlies
the $\pi$ Berry phase associated with the Fermi surface.  Spin resolved
ARPES, described schematically in Fig. \ref{fig:zfig2}(a),
 is ideally suited to probe this physics.
Experiments by \textcite{hsieh09a} measured the spin polarization of the surface
states.  These experiments proved that the surface states are indeed
non degenerate and strongly spin polarized (Fig. \ref{fig:zfig2}(b)),
providing even more decisive
evidence for their topological classification.
In addition, the spin polarization
data also established the connectivity of the
surfaces state bands above $E_F$ (which is inaccessible to ARPES),
showing that bands labeled $2$ and $3$ in Fig. \ref{fig:zfig1}(a) connect to
form a hole pocket.    Finally, they directly mapped
the spin texture of the Fermi surface,
providing the first direct evidence for
the $\pi$ Berry phase by showing that the spin polarization rotates by
$360^\circ$ around the central Fermi surface, shown in
Fig. \ref{fig:zfig2}(c).  The measurement of the {\it handedness} of this
rotation provided even more information about the topological structure,
by probing a {\it mirror Chern number}, which agreed favorably with theory \cite{teofukane08}.

Spin polarized ARPES also enables a similar characterization of
surface states in the metallic regime of
the Bi$_{1-x}$Sb$_x$ series.  Pure Sb is predicted to have a topologically non
trivial valence band, despite the semi metallic band overlap.
\textcite{hsieh09a} found that the surface states of Sb carry
a Berry phase and chirality property predicted by theory \cite{teofukane08}
that is unlike the conventional spin-orbit metals such as gold, which has
zero net Berry phase and no net chirality.
Additional compositions of the Bi$_{1-x}$Sb$_x$ series provided further
evidence for the topological character of the surface states \cite{nishide10}.
These results demonstrate that ARPES and spin-ARPES are powerful probes of topological
order.

\begin{figure}
\includegraphics[width=2.8in]{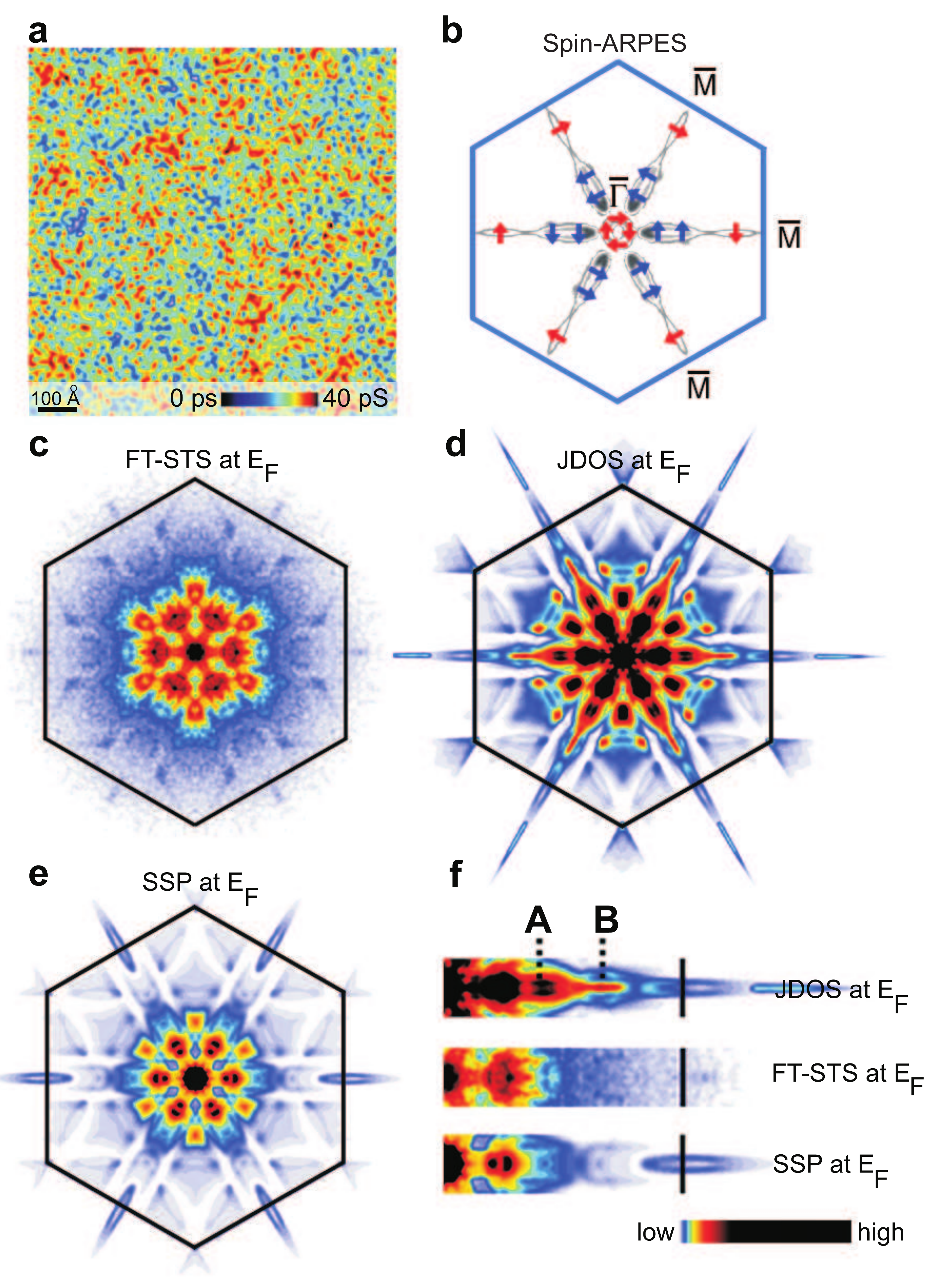}
\caption{ Absence of backscattering:
Quasiparticle interference observed at the surface of Bi$_{0.92}$Sb$_{0.08}$
exhibits and
absence of elastic backscattering: (a) Spatially resolved conductance
maps of the (111) surface obtained at 0 mV
over a 1000\AA$\times$1000\AA. (b) Spin-ARPES map of the
surface state measured at the Fermi level. The spin textures from
spin-ARPES measurements are shown with arrows. (c) Fourier transform
scanning tunneling spectroscopy (FT-STS) at $E_F$. (d) The joint
density of states (JDOS) at $E_F$. (e) The spin-dependent scattering
probability(SSP) at $E_F$. (f) Close-up of the JDOS, FT-STS
and SSP at $E_F$, along the $\Gamma$-M direction.  Adapted from
\onlinecite{hsieh09a,roushan09}.}
\label{fig:zfig3}
\end{figure}

As discussed in section \ref{sec:strongweak}
the topological surface states are expected to be
robust in the presence of non magnetic disorder, and immune from
Anderson localization.  The origin of this is the fact that
${\cal T}$ symmetry forbids the backscattering between Kramers
pairs at ${\bf k}$ and $-{\bf k}$.
Random alloying in Bi$_{1-x}$Sb$_x$,
which is not present in other material families of topological
insulators found to date, makes this material system an ideal
candidate in which to examine the impact of disorder or random
potential on topological surface states. The fact that the
2D states are indeed protected from spin-independent
scattering was established by \textcite{roushan09} by
combining results from scanning tunneling spectroscopy and
spin-ARPES.  Fig. \ref{fig:zfig3} shows
the analysis of the interference pattern
due to scattering at the surface.
Fig. \ref{fig:zfig3}(c) shows the Fourier transform of the observed pattern
(Fig. \ref{fig:zfig3}(a)),
while Figs. \ref{fig:zfig3}(d,e) show the joint density of states computed from
the Fermi surface (Fig. \ref{fig:zfig3}(b))
with and without a suppression of ${\bf k}$ to $-{\bf k}$
backscattering.  The similarity between Figs. \ref{fig:zfig3}(c,e) shows that
despite strong atomic scale disorder,
${\bf k}$ to $-{\bf k}$ backscattering is absent.
Similar conclusions have emerged from studies of the electronic interference
patterns near defects or steps on the surface in other topological insulators
\cite{urazhdin04,zhangt09,alpichshev10}.  In graphene there is an approximate version of
this protection if the disorder has a smooth potential which does
not mix the valleys at ${\bf K}$ and ${\bf K}'$, but real graphene
will become localized with strong disorder \cite{neto09}.

\subsection{Second generation materials: Bi$_2$Se$_3$, Bi$_2$Te$_3$, Sb$_2$Te$_3$}
\label{sec:bise}

The surface structure of Bi$_{1-x}$Sb$_x$ was
rather complicated and the band gap was rather small.  This
motivated a search for topological insulators with a larger band gap
and simpler surface spectrum.
A second generation of 3D
topological insulator materials \cite{moore09}, especially Bi$_2$Se$_3$, offer the
potential for topologically protected behavior in ordinary crystals
at room temperature and zero magnetic field.
In 2008, work led by the Princeton group used ARPES and
first principles calculations to study the surface band structure of
Bi$_2$Se$_3$ and observed the characteristic signature of a
topological insulator in the form of a single Dirac cone \cite{xia09a}.
Concurrent theoretical work by \textcite{zhangh09}
used electronic structure methods to show that
Bi$_2$Se$_3$ is just one of several new large band gap topological
insulators. \textcite{zhangh09} also provided a simple tight-binding
model to capture the single Dirac cone observed in these materials.
Detailed and systematic surface
investigations of Bi$_2$Se$_3$ \cite{hsieh09b,hor09,park10}, Bi$_2$Te$_3$
\cite{chen09,hsieh09b,hsieh09c,xia09b} and  Sb$_2$Te$_3$ \cite{hsieh09c}
confirmed the topological band structure of all 3 of these materials.
This also explained earlier
puzzling observations on Bi$_2$Te$_3$ \cite{noh08}.
These works showed that the topological insulator behavior in
these materials is associated with a band inversion at ${\bf k}=0$, leading to
the $(1;000)$ topological class.
The $(1;000)$ phase observed in the Bi$_2$Se$_3$ series
differs from the $(1;111)$ phase in Bi$_{1-x}$Sb$_x$ due to its
weak topological invariant, which has implications for the behavior
of dislocations\cite{ran09}.

\begin{figure}
\includegraphics[width=3in]{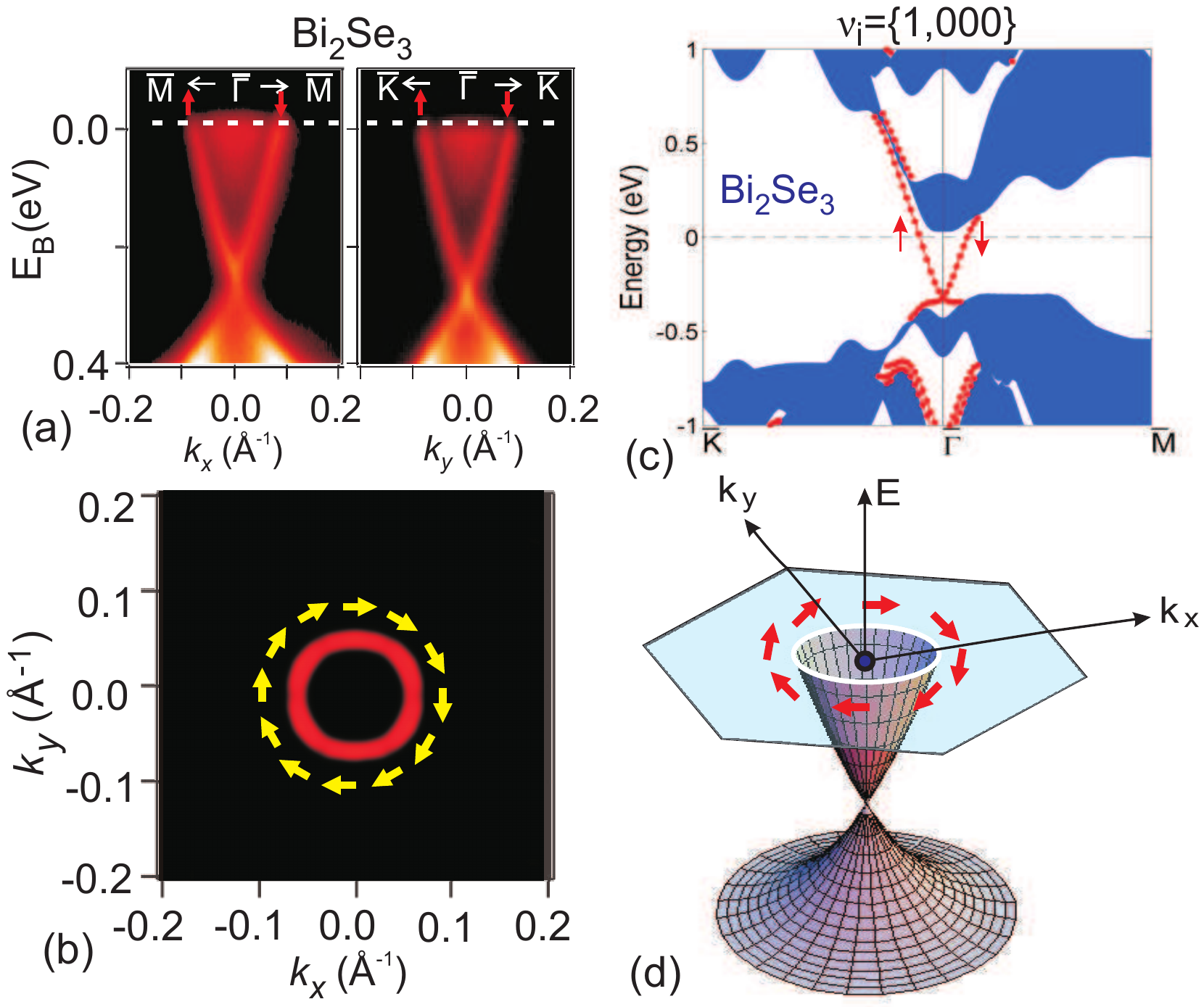}
\caption{\label{fig:zfig4}
Helical fermions: Spin-momentum
locked helical surface Dirac fermions are hallmark signatures of
topological insulators.  (a) ARPES data for Bi$_2$Se$_3$ reveals
surface electronic states with a single
spin-polarized Dirac cone. The Surface Fermi surface (b) exhibits a chiral
left-handed spin texture.  (c) Surface electronic structure of
Bi$_2$Se$_3$ computed in the local density approximation.  The shaded
regions describe bulk states, and the red lines are surface states.
(d) Schematic of the spin polarized surface state dispersion in
Bi$_2$X$_3$ $(1;000)$ topological insulators.  Adapted from \onlinecite{hsieh09b,xia08,xia09b}.}
\end{figure}

\begin{figure}
\includegraphics[width=3in]{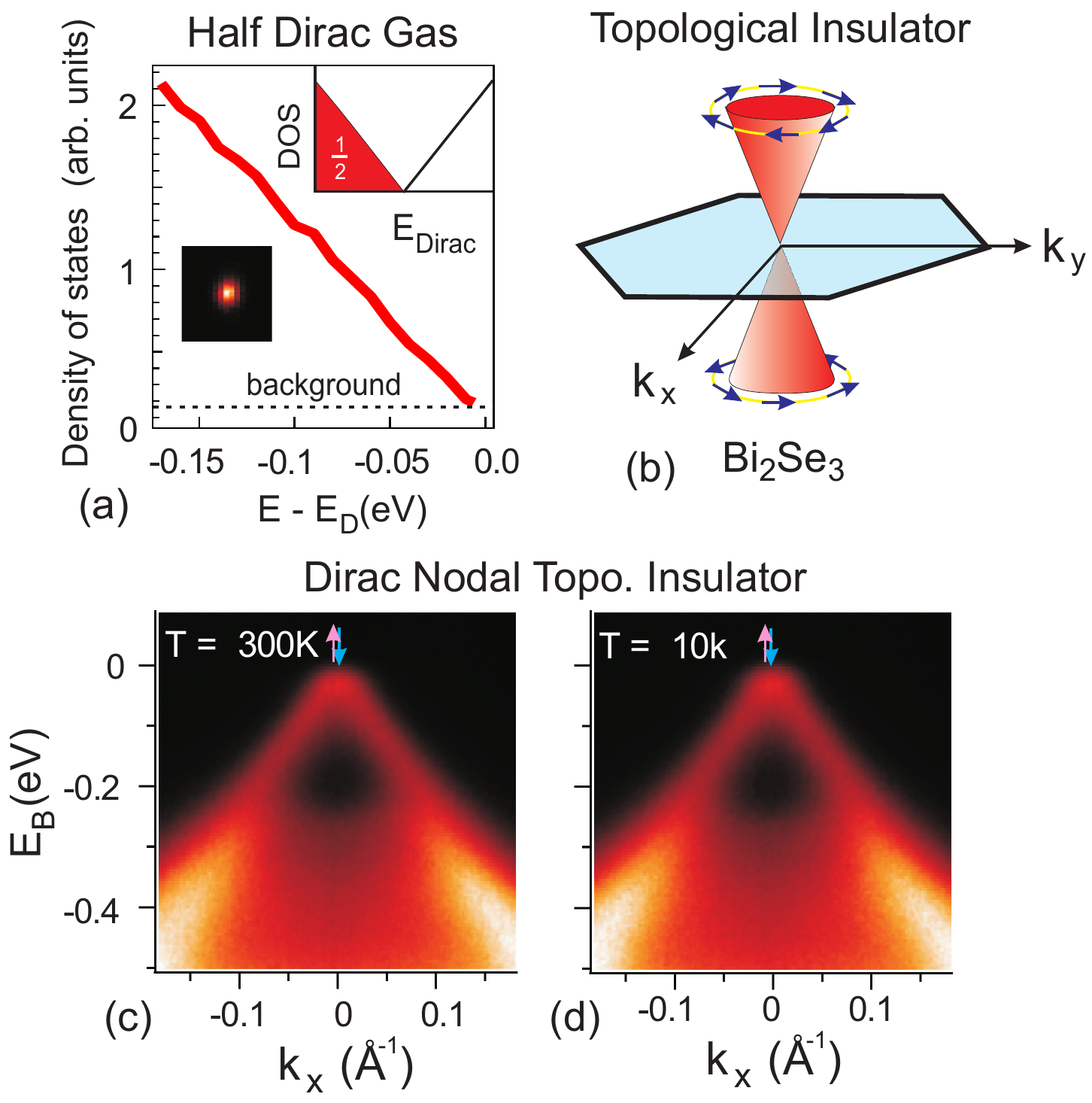}
\caption{ Room temperature topological
order in Bi$_2$Se$_3$:  (a) Crystal momentum
integrated ARPES data near Fermi level exhibit linear fall-off of
density of states, which combined with the spin-resolved nature of
the states suggest that a half Fermi gas is realized on the
topological surfaces. (b) Spin-texture map based on spin-ARPES data
suggest that the spin-chirality changes sign across the Dirac point.
(c) The Dirac node remains well
defined up a temperature of 300K suggesting the stability of
topological effects up to the room temperature.  Adapted from \onlinecite{hsieh09b}.}
\label{fig:zfig7}
\end{figure}

\begin{figure}
\includegraphics[width=3in]{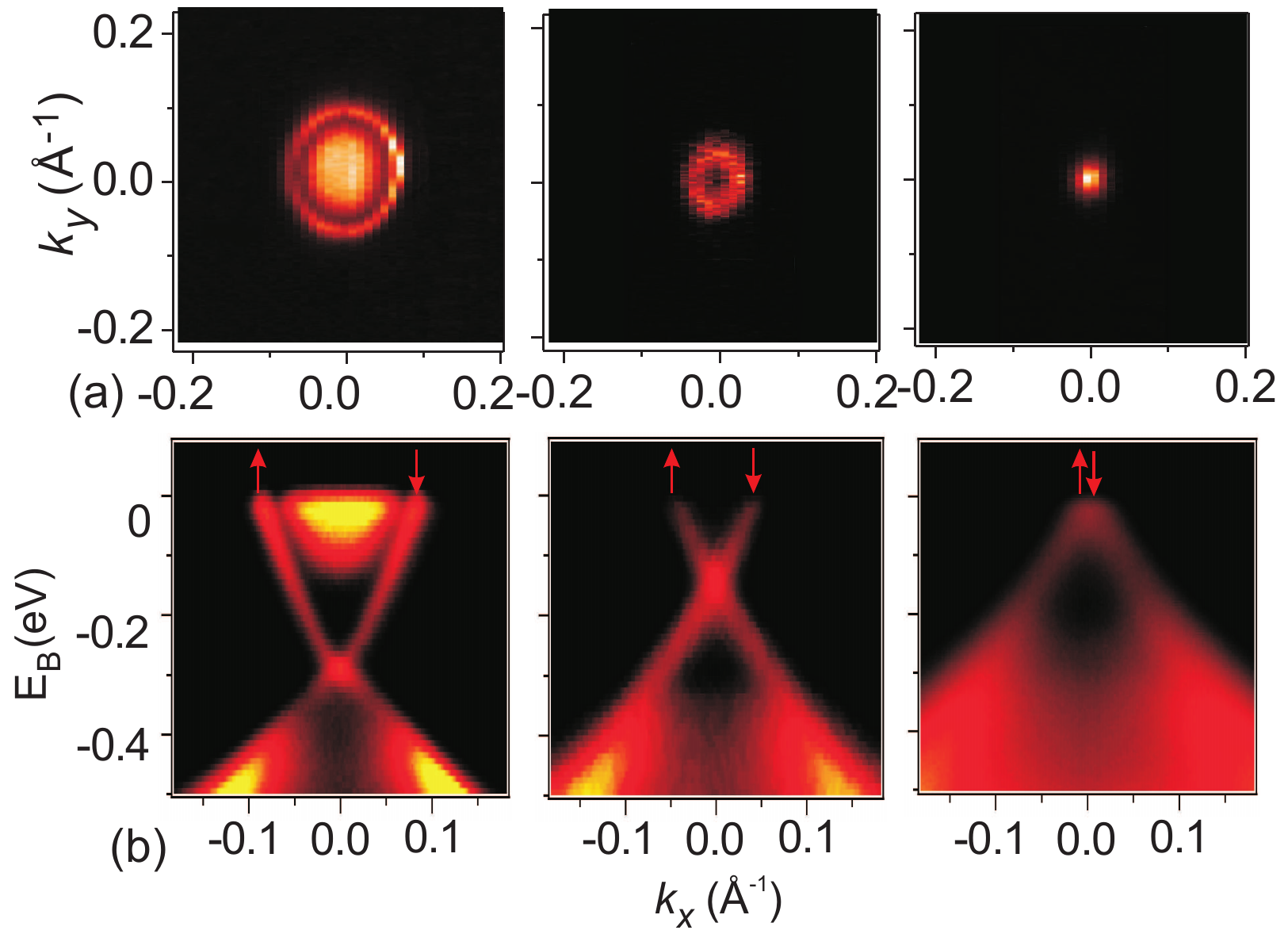}
\caption{
Chemical gating a topological
surface to the spin-degenerate point: Topological insulator surfaces
are most interesting if the chemical potential can be placed at the
Dirac node without intercepting any bulk band. This can be
achieved in Bi$_2$Se$_3$ via the chemical tailoring of the surface or
using electrical gating methods. (a) Evolution of surface Fermi
surface with increasing NO$_2$ adsorption on the surface.
NO$_2$ extracts electrons from the Bi$_2$Se$_3$ surface leading to an
effective hole doping of the material. (b) Chemical gating of the
surface can be used to place the chemical potential at the spin
degenerate Dirac point.  Adapted from \onlinecite{hsieh09b}.}
\label{fig:zfig6}
\end{figure}

Though the phase observed in the Bi$_2$Se$_3$ class has the same
strong topological invariant $\nu_0=1$ as Bi$_{1-x}$Sb$_x$,
there are three
crucial differences that suggest that this series may become the
reference material for future experiments. The
Bi$_2$Se$_3$ surface state is found from ARPES and theory to be
a nearly idealized single Dirac cone as seen from the experimental
data in Figs. \ref{fig:zfig4},\ref{fig:zfig7},\ref{fig:zfig6}.
Second, Bi$_2$Se$_3$ is stoichiometric (i.e., a pure
compound rather than an alloy like Bi$_{1-x}$Sb$_x$) and hence can be
prepared in principle at higher purity.  While the topological
insulator phase is predicted to be quite robust to disorder, many
experimental probes of the phase, including ARPES of the surface band
structure, are clearer in high-purity samples. Finally, and perhaps
most important for applications, Bi$_2$Se$_3$ has a large band gap of
approximately 0.3 eV (3600$^\circ$K).  This indicates that in its high
purity form Bi$_2$Se$_3$ can exhibit topological insulator behavior at
{\it room temperature}(Fig. \ref{fig:zfig7})
and greatly increases the
potential for applications. To understand the likely impact
of these new topological insulators, an analogy can be drawn with the
early days of high-temperature cuprate superconductivity: the
original cuprate superconductor LBCO was quickly superseded by
``second-generation" materials such as YBCO and BSCCO for most applied
and scientific purposes.

\begin{figure}
\includegraphics[width=3in]{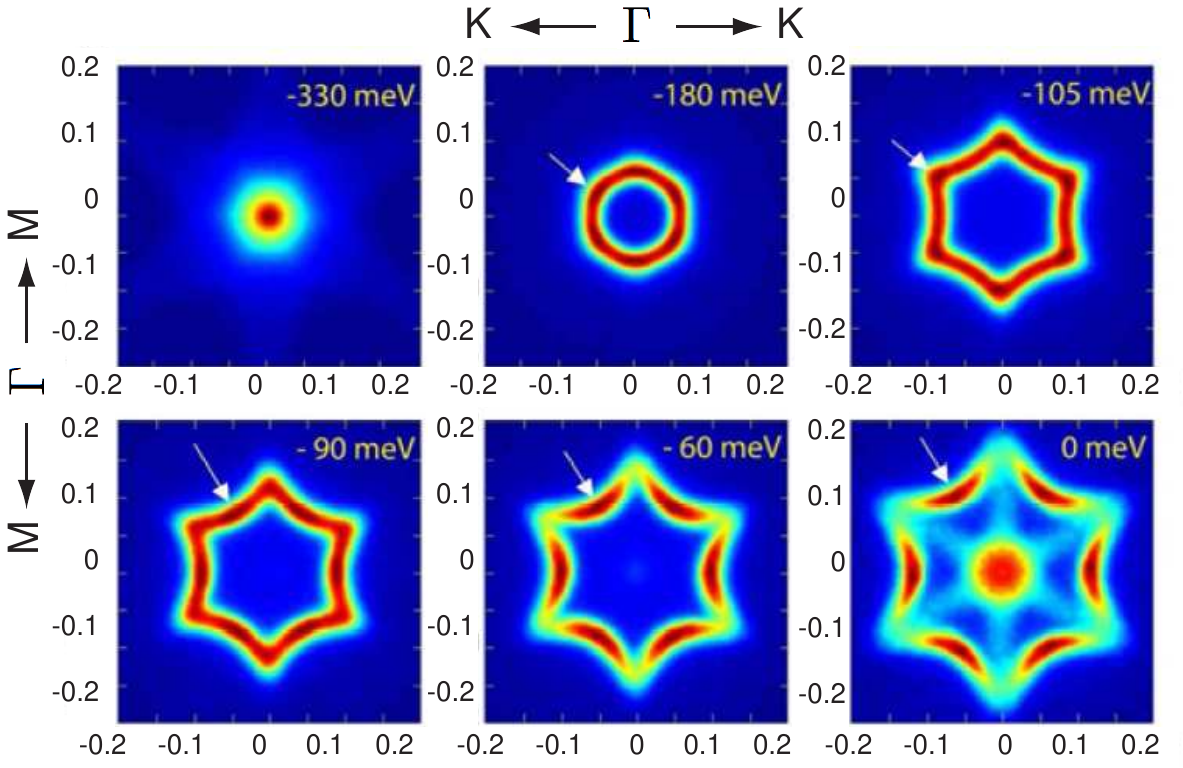}
\caption{Hexagonal warping of
surface states in Bi$_2$Te$_3$: ARPES and STM studies of Bi$_2$Te$_3$
reveal a hexagonal deformation of surface states.  Fermi surface evolution with
increasing n-type doping as observed in ARPES measurements.  Adapted from
\onlinecite{alpichshev10}.}
\label{fig:zfig8}
\end{figure}

All the key properties of
topological states have been demonstrated for Bi$_2$Se$_3$ which has
the simplest Dirac cone surface spectrum and the largest band gap.
In Bi$_2$Te$_3$ the surface states exhibit large deviations
from a simple Dirac cone (Fig. \ref{fig:zfig8}) due to a combination of smaller
band gap (0.15 eV) and a strong trigonal potential \cite{chen09},
which can be utilized to explore some aspects
of its surface properties \cite{fu09,hasan09}.
The hexagonal deformation of the surface states is confirmed by STM
measurements \cite{alpichshev10} (Fig. \ref{fig:zfig8}).
Speaking of applications within this class of
materials, Bi$_2$Te$_3$, is already well known to materials
scientists working on thermoelectricity.  It is a commonly used
thermoelectric material in the crucial engineering regime near room
temperature.

\begin{figure}
\includegraphics[width=3in]{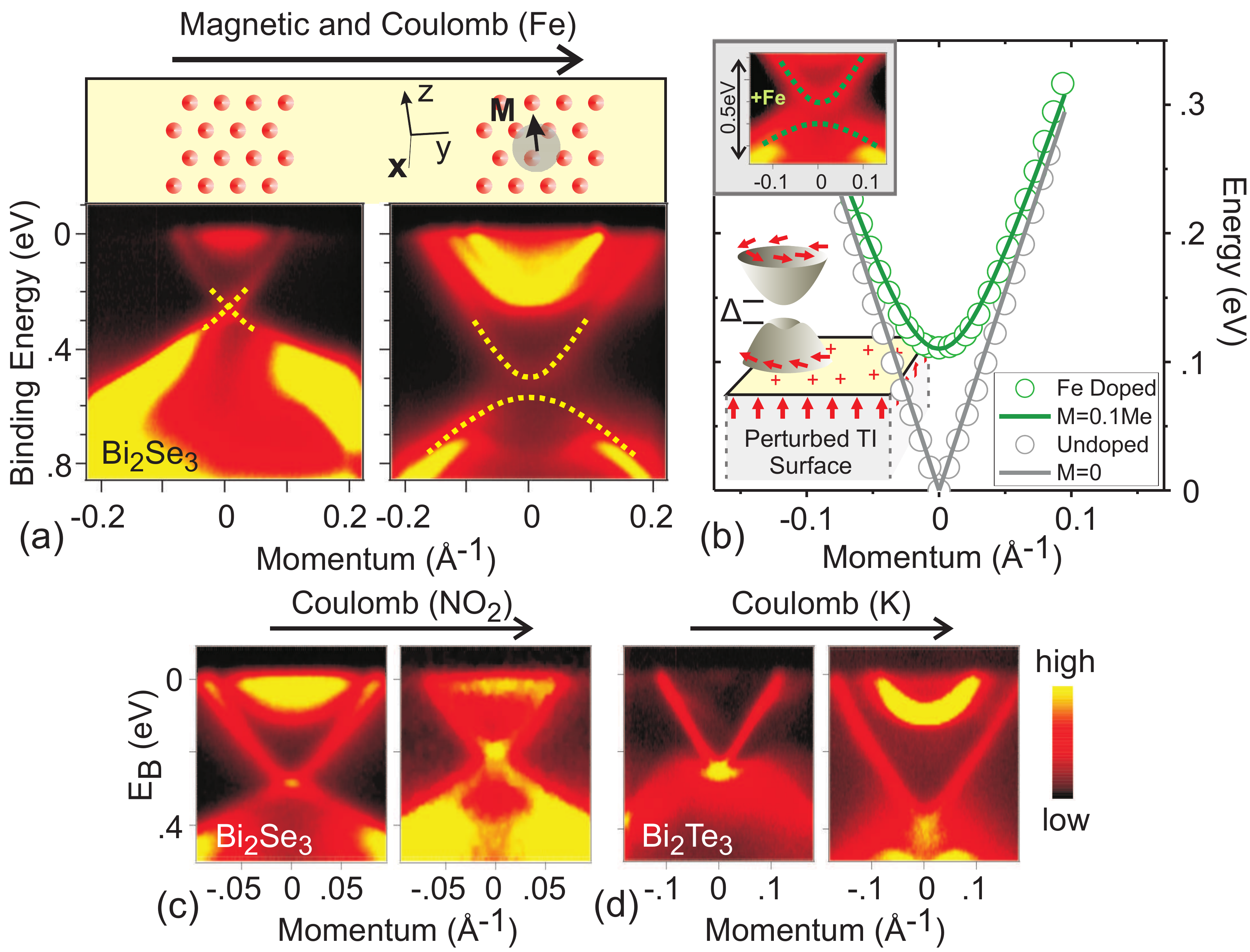}
\caption{ Protection by time reversal
symmetry: Topological surface states are robust in the presence of
strong non-magnetic disorder but open a gap in the presence of
${\cal T}$ breaking magnetic impurities and disorder. (a) Magnetic
impurity such as Fe on the surface of Bi$_2$Se$_3$ opens a gap at the
Dirac point. The magnitude of the gap is set by the interaction of Fe
ions with the Se surface and the ${\cal T}$ breaking disorder
potential introduced on the surface. (b) A comparison of surface band
dispersion with and without Fe doping. (c,d) Non-magnetic disorder
created via molecular absorbent NO$_2$ or alkali atom adsorption (K
or Na) on the surface leaves the Dirac node intact in both
Bi$_2$Se$_3$ and Bi$_2$Te$_3$.  Adapted from \onlinecite{hsieh09b,xia09b,wray10}.}
\label{fig:zfig5}
\end{figure}

Two defining properties of topological insulators --
spin-momentum locking of surface states and $\pi$ Berry phase -- can
be clearly demonstrated in the Bi$_2$Se$_3$ series.
The surface states are expected to be protected by ${\cal T}$ symmetry
which implies that the surface Dirac node should be robust in
the presence of non-magnetic disorder but open a gap in the presence
of ${\cal T}$ breaking perturbations.
Magnetic impurities such as Fe or Mn on the surface of Bi$_2$Se$_3$ open a gap at the
Dirac point (Fig. \ref{fig:zfig5}(a,b)) \cite{xia08,hsieh09b,hor10b,wray10}.
The magnitude of the gap is
likely set by the interaction of Fe ions with the Se surface and the
${\cal T}$ breaking disorder potential introduced on the surface.
Non-magnetic disorder created via molecular absorbent NO$_2$ or
alkali atom adsorption (K or Na) on the surface leaves the Dirac node
intact (Fig. \ref{fig:zfig5}(c,d)) in both Bi$_2$Se$_3$ and Bi$_2$Te$_3$
\cite{xia09b,hsieh09b}. These results
are consistent with the fact that the topological surface states are
protected by ${\cal T}$ symmetry.

Many of the interesting theoretical proposals that utilize
topological insulator surfaces
require the chemical potential to lie at or near the surface Dirac point.
This is similar to the case in graphene, where
the chemistry of carbon atoms naturally locates the Fermi
level at the Dirac point.   This makes its
density of carriers highly tunable by an applied electrical field and
enables applications of graphene to both basic science and
microelectronics.  The surface Fermi level of a topological insulator
depends on the detailed electrostatics of the surface, and is not necessarily
at the Dirac point.  Moreover, for naturally grown Bi$_2$Se$_3$ the {\it bulk} Fermi
energy is not even in the gap.  The observed $n$ type behavior is believed to
be caused Se vacancies.  By appropriate chemical modifications, however, the
Fermi energy of both the bulk and the surface can be controlled.
This allowed \textcite{hsieh09b} to reach the sweet spot in which the surface
Fermi energy is tuned to the Dirac point (Fig. \ref{fig:zfig6}).  This
was achieved by doping bulk
with a small concentration of Ca,
which compensates the Se vacancies, to place the Fermi level within
the bulk band gap.  The surface was hole doped by exposing the
surface to NO$_2$ gas to place the Fermi level at the Dirac point.

\begin{figure}
\includegraphics[width=3in]{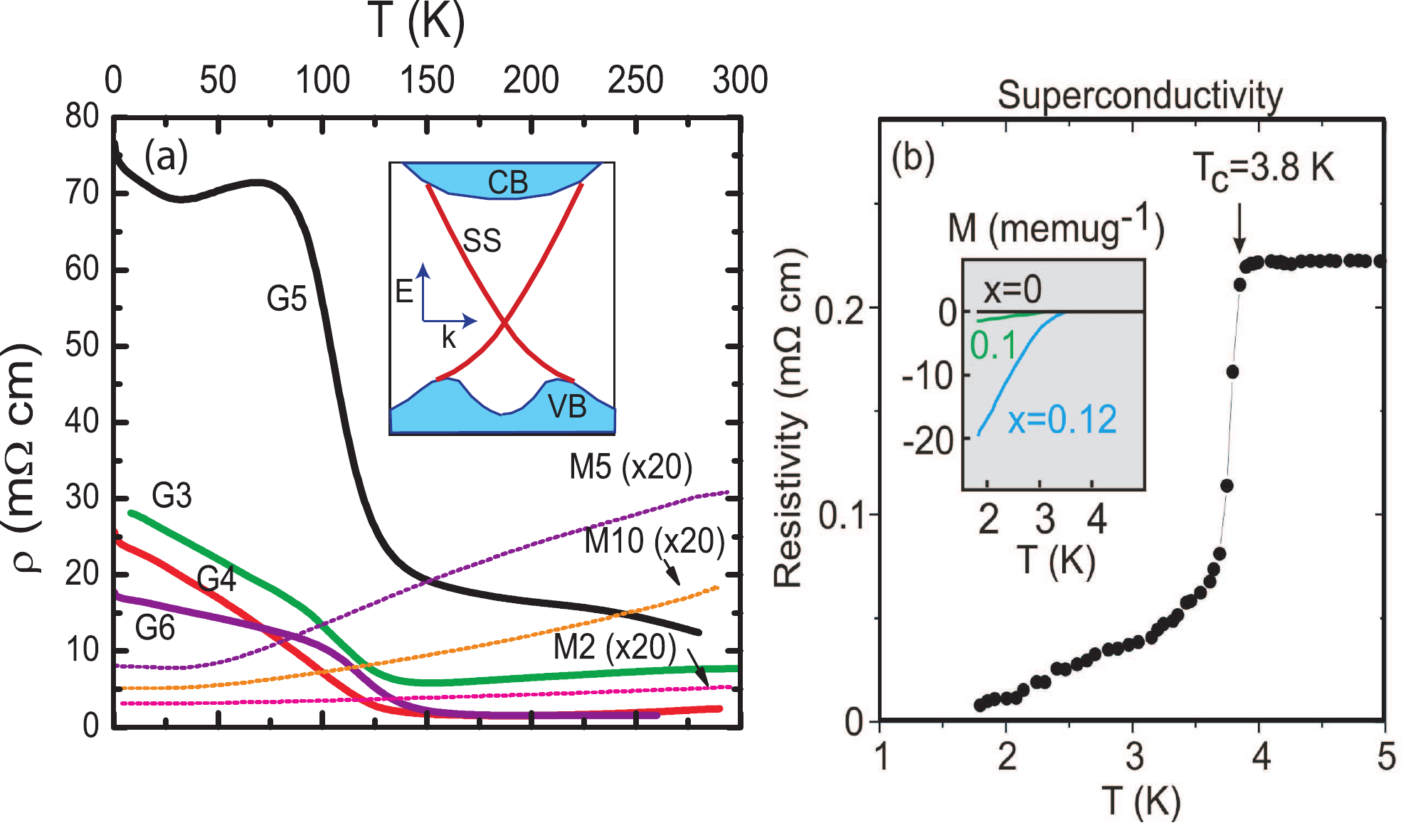}
\caption{Electrical transport in Bi$_2$Se$_3$.  (a) Resistivity for samples of
pure Bi$_2$Se$_3$ doped with a small concentration of Ca.  Increasing the Ca
concentration moves the Fermi level from the conduction band into the gap and
then to the valence band.  Samples with $.002<x<.0025$, labeled G,
show insulating behavior below 100$^\circ$K \cite{checkelsky09}.  (b) Bi$_2$Se$_3$ doped with Cu shows superconducting behavior
below 3.8$^\circ$K for $x = .12$.  The inset shows the magnetic susceptibility
which exhibits the Meissner effect.  Adapted from \onlinecite{hor10a,wray10}.
}
\label{fig:transport}
\end{figure}

The main remaining complication with these materials, especially for
experimental techniques that (unlike ARPES) do not distinguish
directly between bulk and surface states, is that they have some
residual conduction in the bulk from impurity or self doping states.
Electrical transport measurements on Bi$_2$Se$_3$ show that doping with a small
concentration of Ca leads to insulating behavior.  Fig. \ref{fig:transport}(a)
shows the resistivity of several samples with varying Ca concentrations.  For
$.002<x<.025$, the resistivity shows a sharp upturn below 100$^\circ$K before
saturating.  The low temperature resistivity is still too small to be explained by
the surface states alone.  However, the low temperature transport exhibits
interesting 2D mesoscopic effects that are not completely
understood \cite{checkelsky09}.   Doping Bi$_2$Se$_3$ with copper leads to a
metallic state that shows superconducting behavior (Fig.
\ref{fig:transport}(b)) below 3.8$^\circ$K \cite{hor10a,wray09}.  This has important
ramifications for some of the devices proposed in the following section.

\section{Exotic Broken Symmetry Surface Phases}
\label{sec:exoticsurface}

Now that the basic properties of topological insulators have been established,
we may ask what can be done with them.  In this section we will argue
that the unique properties of topological insulator surface and edge
states are most dramatic if an energy gap can be induced in them.
This can be done by breaking ${\cal T}$ symmetry with an external
magnetic field \cite{fukane07} or proximity to a magnetic material \cite{qihugheszhang08}, by breaking gauge
symmetry due to proximity to a superconductor \cite{fukane08}, or by an excitonic
instability of two coupled surfaces \cite{seradjeh09}.  In this section we review
the magnetic and superconducting surface phases.

\subsection{Quantum Hall effect and topological magnetoelectric effect}
\label{sec:qhetopomag}

\subsubsection{Surface quantum Hall effect}
\label{sec:surfaceqhe}

A perpendicular magnetic field will lead to Landau levels in the
surface electronic spectrum, and the quantum Hall effect.
The Landau levels for Dirac electrons
are special, however, because a Landau level is guaranteed to exist
at exactly zero energy \cite{jackiw84}.  This zero Landau level is particle-hole symmetric
in the sense that the Hall conductivity is equal and opposite when the
Landau level is full or empty.  Since the Hall conductivity increases by
$e^2/h$ when the Fermi energy crosses a Landau level
the Hall conductivity is {\it half integer} quantized \cite{ando02},
\begin{equation}
\sigma_{xy} = (n+1/2) e^2/h.
\label{n+1/2}
\end{equation}

\begin{figure}
\includegraphics[width=2.8in]{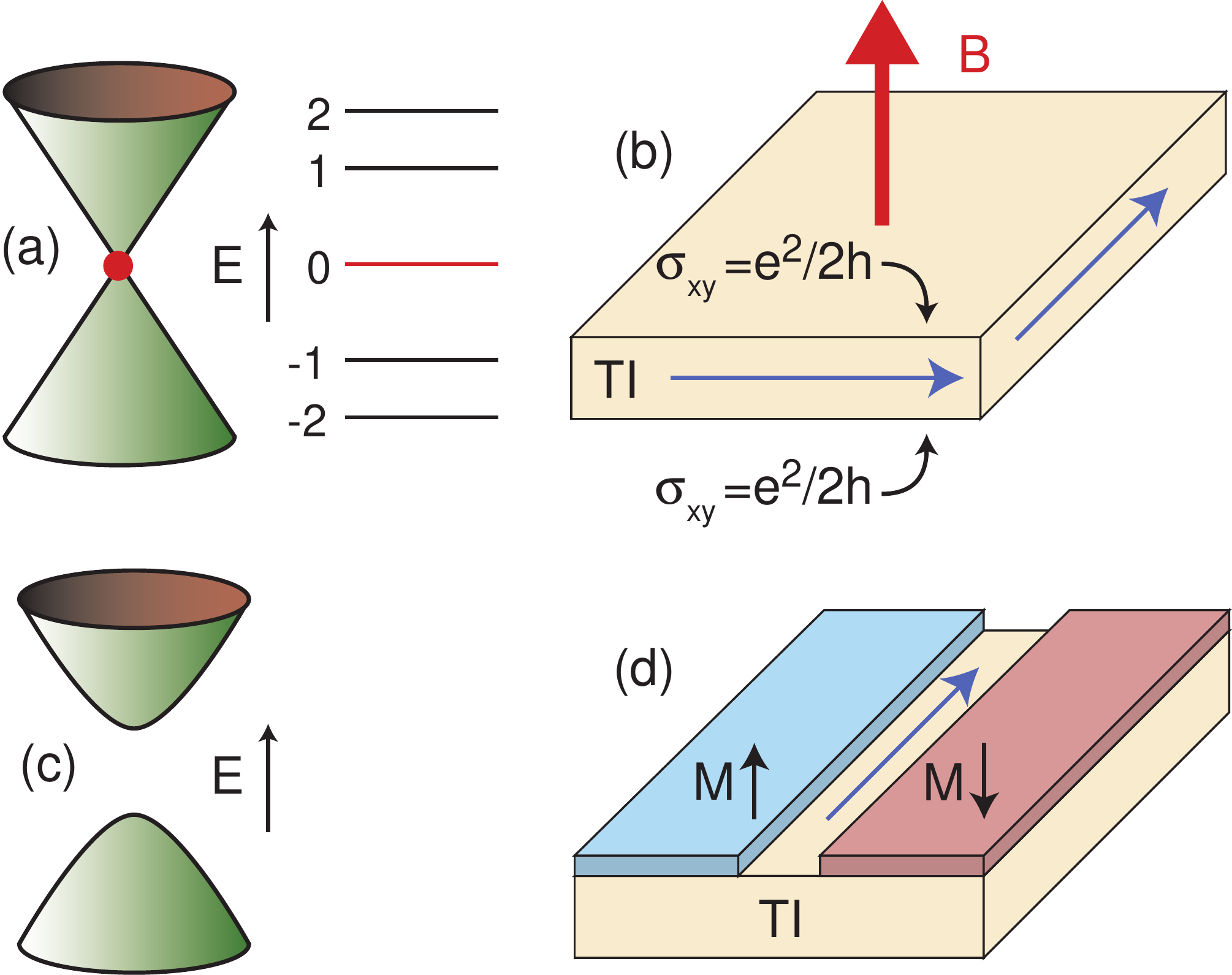}
\caption{Surface quantum Hall effect.  (a) The Dirac spectrum is replaced by Landau levels
in an orbital magnetic field.  (b) The top and bottom surfaces share a single chiral fermion
edge mode.  (c) A thin magnetic film can induce an energy gap at the surface.
(d) A domain wall in the surface magnetization then exhibits a chiral fermion mode.}
\label{fig:surfaceqhe}
\end{figure}

This physics has been famously demonstrated in experiments on
graphene \cite{novoselov05,zhangy05}.  However, there is an important difference.  In graphene
\eqref{n+1/2} is multiplied by four, due to the spin and valley degeneracy
of graphene's Dirac points, so the observed Hall conductivity is
still integer quantized.  At the surface of the topological
insulator there is only a {\it single} Dirac point.  Such a
``fractional" integer quantized Hall effect should be a cause for
concern because the integer quantized Hall effect is always
associated with chiral edge states, that can only be integer
quantized.  The resolution is the mathematical fact
that a surface can not have
a boundary.  In a slab geometry shown in Fig. \ref{fig:surfaceqhe}(a), the
top surface and bottom surface are necessarily connected to each other, and will
always be measured in parallel \cite{fukane07}, doubling the $1/2$.  The top and
bottom can share a single chiral edge state, which carries the integer
quantized Hall current.

A related surface quantum Hall effect, called the anomalous quantum Hall
effect, can be induced with the proximity to a
magnetic insulator.  A thin magnetic
film on the surface of a topological insulator will give rise to a local exchange
field that lifts the Kramers degeneracy at the surface Dirac points.  This
introduces a mass term $m$ into the Dirac equation \eqref{surfacedirac},
as in \eqref{diracmass}.  If the $E_F$ is in this energy
gap, there is a half integer quantized Hall conductivity
$\sigma_{xy}=e^2/2h$\cite{pankratov87}, as discussed in section \ref{sec:dirac}.
This can be probed in a transport experiment by introducing a {\it domain
wall} into the magnet.  The sign of $m$ depends on the direction of
the magnetization.  At an interface where $m$ changes sign (Fig. \ref{fig:surfaceqhe}(d))
there will be a 1D chiral edge state, analogous to unfolding the
surface in Fig. \ref{fig:surfaceqhe}(b).

\subsubsection{Topological magnetoelectric effect and axion electrodynamics}
\label{sec:topomag}

The surface Hall conductivity can also be probed without the
edge states either by optical methods or by measuring the magnetic field produced by surface
currents.  This leads to an intriguing {\it topological magnetoelectric effect} \cite{qihugheszhang08,essin09}.
Imagine a cylindrical topological insulator with magnetically gapped surface
states and an electric field ${\bf E}$ along its axis.  The azimuthal surface Hall
current $(e^2/2h)|{\bf E}|$ leads to a magnetic dipole moment associated
with a magnetization ${\bf M} = \alpha{\bf E}$, where the magnetoelectric polarizability
is given by $\alpha = e^2/2h$.

A field theory for this magnetoelectric effect can be developed by including a
``$\theta$ term" in the electromagnetic Lagrangian, which has a form analogous to
the theory of axion electrodynamics that has been studied in particle
physics contexts \cite{wilczek87},
\begin{equation}
\Delta {\cal L} = \theta (e^2/2\pi h) {\bf E}\cdot{\bf B}.
\label{edotb}
\end{equation}
The field $\theta$, which is a dynamical variable in the axion theory, is
a constant, $\pi$, in the topological insulator.
Importantly, when expressed in terms of the vector potential ${\bf E}\cdot {\bf
B}$ is a total derivative, so a constant $\theta$ has no effect on the
electrodynamics.  However, a gapped interface,
across which $\theta$ changes by $\Delta\theta$, is
associated with a surface Hall conductivity $\sigma_{xy} = \Delta\theta e^2/(2\pi h)$.

As in the axion theory, the action corresponding to \eqref{edotb} is invariant under
$\theta \rightarrow \theta+2\pi$.  Physically, this reflects the fact
that an integer quantum Hall state with $\sigma_{xy}=ne^2/h$
can exist at the surface without changing the bulk properties \cite{essin09}.
This resembles a similar ambiguity in the
{\it electric polarization}.
\textcite{qihugheszhang08} showed that
since ${\bf E}\cdot{\bf B}$ is odd under ${\cal T}$,
only $\theta=0$ or $\pi$ are consistent with ${\cal T}$ symmetry,
so $\theta$ is quantized.
By computing the magnetoelectric response perturbatively,
$\theta$ can be computed in a
manner similar to the Kubo formula calculation of $\sigma_{xy}$.
$\theta/\pi$ is identical to $\nu_0$, the invariant
characterizing a strong topological insulator.

Observation of the surface currents associated with this magnetoelectric effect
will be an important complement to the ARPES experiments.  It should be
emphasized, however, that despite the topologically quantized status of
$\theta$, the surface currents are not quantized the way edge state {\it transport} currents
are quantized in the quantum Hall effect.  The
surface currents are {\it bound} currents, which must be distinguished from
other bound currents that may be present.  Nonetheless, it may be possible to
account for such effects, and signatures of $\theta$
will be interesting to observe.  \textcite{qizhang09} pointed out that a
consequence of a nonzero surface $\sigma_{xy}$ is that an electric
charge outside the surface gives rise to a pattern of surface
currents that produces a magnetic field the same as that of an image
magnetic monopole.

\subsection{Superconducting proximity effect}
\label{sec:proximity}

Combining topological insulators with ordinary superconductors leads
to an exquisitely correlated interface state that, like a topological
superconductor, is predicted to host Majorana fermion excitations.
In this section we will begin by reviewing the properties of Majorana
fermion excitations and the ingenious proposal by \textcite{kitaev03} to use those
properties for fault tolerant quantum information processing.  We
will then describe methods for engineering Majorana fermions in
superconductor-topological insulator devices and prospects for their
experimental observation.

\subsubsection{Majorana fermions and topological quantum computing}
\label{sec:qcompute}

As discussed in section \ref{sec:majoranaedge}, a well separated pair of
Majorana bound states
defines a degenerate two level system -- a qubit.  Importantly, the quantum
information in the qubit is stored non locally.  The state
can not be measured with a local measurement on one of the bound states.  This
is crucial, because the main difficulty with making a quantum computer is
preventing the system from accidentally measuring itself.  $2N$ Majorana bound states
defines $N$ qubits -- a quantum memory.

\begin{figure}
\includegraphics[width=2in]{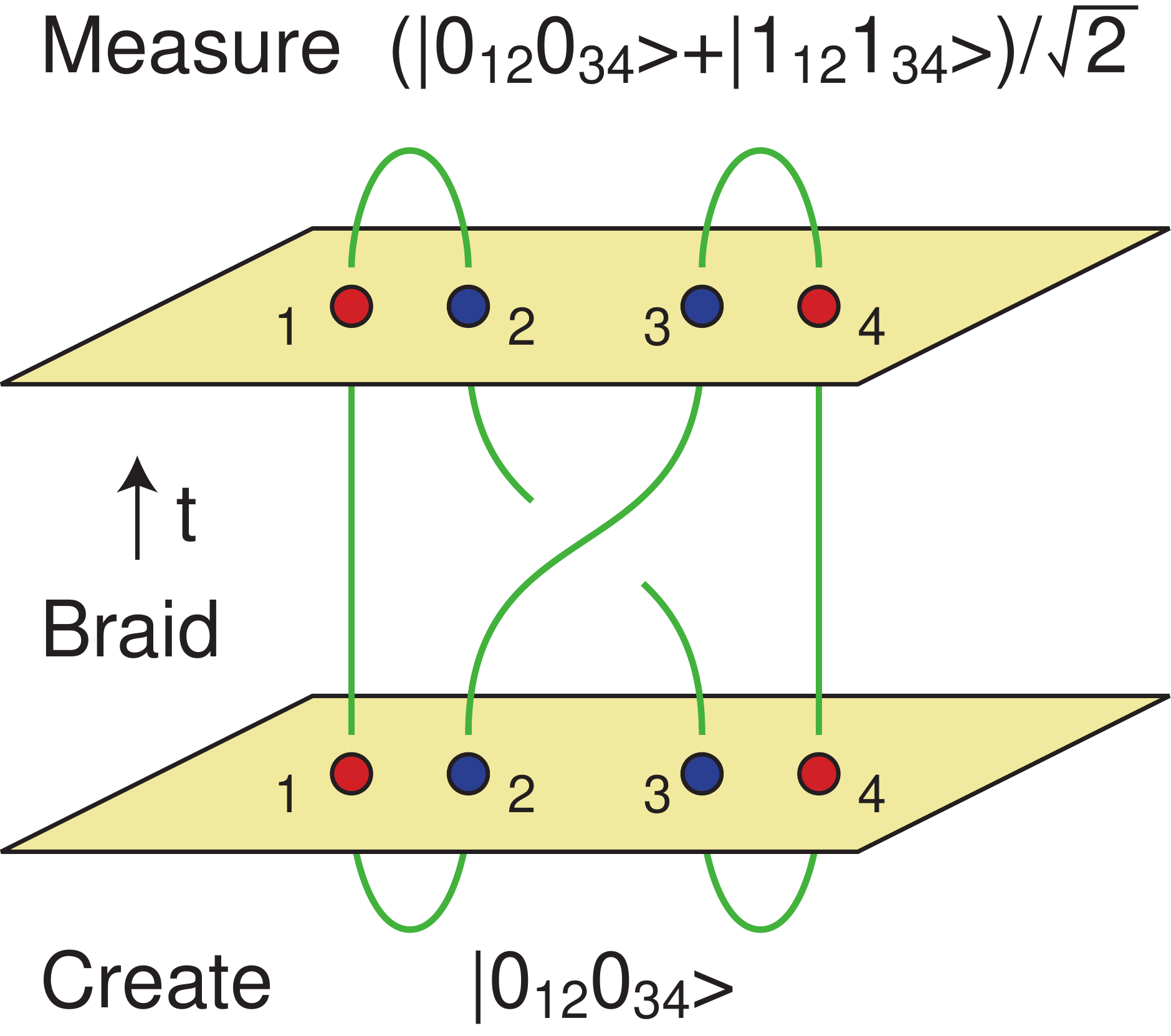}
\caption{A simple operation in which two vortices are exchanged.  The vortex pairs 12 and 34 are
created in the vacuum (0 quasiparticle) state.
When they are brought back together they are in an entangled superposition of
0 and 1 quasiparticle states.}
\label{fig:braid}
\end{figure}

Adiabatically interchanging the vortices, or more generally braiding them, leads to
the phenomenon of {\it non-Abelian statistics} \cite{mooreread91}.  Such processes implement unitary
operations on the state vector $|\psi_a\rangle\rightarrow
U_{ab}|\psi_b\rangle$ that generalize the usual notion of Fermi and Bose quantum
statistics \cite{nayak96,ivanov01}.  These operations are precisely
what a quantum computer is supposed to do.
A quantum computation will consist of three steps, depicted in Fig. \ref{fig:braid}:

{\it (i) Create}:  If a pair $i, j$ of
vortices is created, they will be in the ground state $|0_{ij}\rangle$
with no extra quasiparticle excitations.  Creating $N$ pairs
initializes the system.

{\it (ii) Braid}:  Adiabatically rearranging
the vortices modifies the state, and performs a quantum computation.

{\it (iii) Measure}: Bringing vortices $i$ and $j$ back together allows
the quantum state associated with each pair to be measured.
$|1_{ij}\rangle$ and $|0_{ij}\rangle$ will be distinguished by the presence or
absence of an extra fermionic quasiparticle associated with the pair.

Though the quantum operations allowed by manipulating the Majorana states
do not have sufficient structure to construct a universal quantum
computer \cite{freedman02}, the topological protection of the quantum information
makes the experimental observation of Majorana fermions and non-Abelian
statistics a high priority in condensed matter
physics \cite{nayak08}.   Current experimental efforts have focused on the
$\nu=5/2$ quantum Hall state, where interferometry experiments \cite{sternhalperin06,dassarma05}  can
in principle detect the non-Abelian statistics predicted for the quasiparticles.
Though recent experiments on the quantum Hall effect
have shown encouraging indirect evidence for these states \cite{dolev08,radu08,willett09},
definitive observation of the Majorana states has remained elusive.
In the following section we will describe the possibility of
realizing these states in topological insulator-superconductor
structures.   The large energy scale associated with the energy gap
in Bi$_2$Se$_3$ may provide an advantage, so the required temperature
scale will be limited only by the superconductor.

\subsubsection{Majorana fermions on topological insulators}
\label{sec:majoranaontopo}

\begin{figure}
\includegraphics[width=3in]{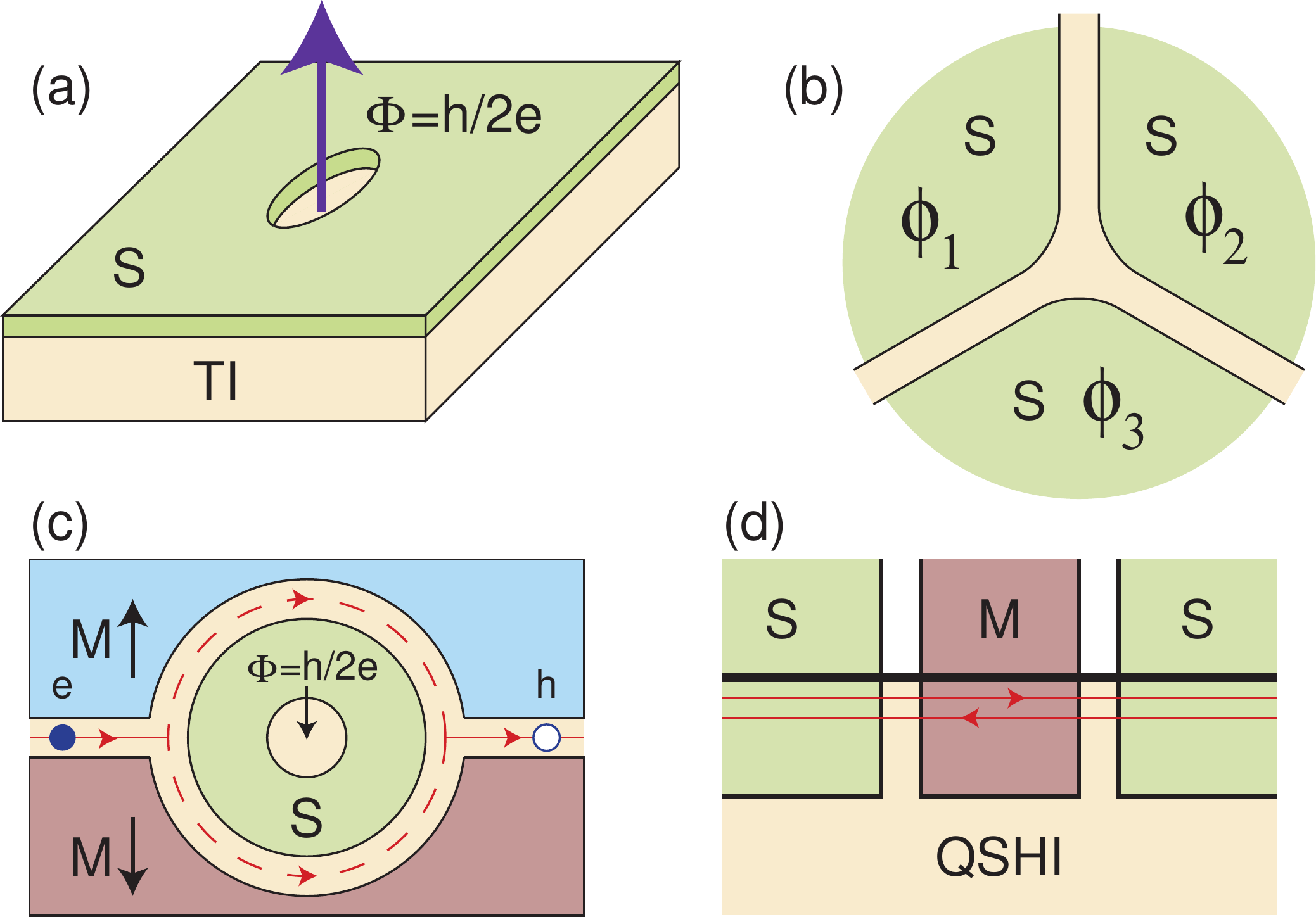}
\caption{Majorana fermions on topological insulators.  (a) A superconducting vortex, or antidot
with flux $h/2e$ on a topological insulator is associated with a Majorana zero mode.  (b)
a superconducting tri-junction on a topological insulator.  Majorana modes
at the junction can be controlled by adjusting the phases $\phi_{1,2,3}$.  1D chiral
Majorana modes exist at a superconductor-magnet interface on a topological
insulator. (c)
shows a 1D chiral Dirac mode on a magnetic domain wall
that splits into two chiral Majorana modes around a superconducting island.
When $\Phi = h/2e$ interference of the Majorana modes converts an electron
into a hole.  (d) Majorana modes at a superconductor-magnet junction on a 2D QSHI.
}
\label{fig:majorana}
\end{figure}

Consider an interface between a topological insulator and an $s$ wave superconductor.
Due to the superconducting proximity effect, Cooper pairs may tunnel from the
superconductor to the surface, leading to
an induced superconducting energy gap in the surface states.  The resulting
2D superconducting state is different from an ordinary
superconductor because the surface states are not spin degenerate and
contain only half the degrees of freedom of a normal metal.  The
superconducting state resembles the spinless $p_x+i p_y$ topological superconductor discussed in
section \ref{sec:superconductor}, which is also based on a spin non degenerate Fermi
surface.   Unlike the $p_x+i p_y$ superconductor, the surface superconductor does not
violate ${\cal T}$ symmetry, and its Cooper pairs have even
parity.  The minus sign required by Fermi statistics is supplied by the $\pi$
Berry phase of the surface states.  Like the $p_x+ip_y$ superconductor,
the surface superconductor will have a zero energy Majorana state
bound to a vortex \cite{fukane08}.
Similar zero modes were later found for superconducting
graphene\cite{ghaemi07,bergman09}, though those modes were intrinsically doubled.
Undoubled Majorana bound states were found earlier by \textcite{jackiw81}
in a related field theory model that had an extra chiral symmetry.
Interestingly, the Majorana states on a topological insulator
emerge as solutions to a 3D BdG theory, so there is a sense in which their
non-Abelian statistics is inherently three dimensional \cite{teokane10}.

Majorana states can in principle be engineered and manipulated by using junctions of superconductors
on the surface of a topological insulator \cite{fukane08}.   If the phases on three superconductors
that meet at a tri-junction (Fig. \ref{fig:majorana}(b)) are arranged such that $(\phi_1,\phi_2,\phi_3) =
(0,2\pi/3,4\pi/3)$, then a vortex is simulated, and a zero mode will
be bound to the junction.  If the phases are changed, the zero mode
can not disappear until the energy gap along one of the three linear
junctions goes to zero.  This occurs when the phase difference across
the junction is $\pi$.  At this point the Majorana bound state moves to the
other end of the linear junction.  Combining these tri-junctions into
circuits connected by linear junctions could then allow for
the {\it Create-Braid-Measure} protocol discussed in section
\ref{sec:qcompute} to be implemented.
The state of two Majorana modes brought
together on a linear junction can be probed by measuring the
supercurrent across that junction.

There are many hurtles to overcome before this ambitious proposal can be realized.
The first step is finding a suitable superconductor that makes good
contact with a topological insulator.  Probing the signatures of
Majorana fermions and non-Abelian statistics will require ingenuity
-- what makes them good for quantum computing makes them
hard to measure.  A first step would be to detect the
Majorana state at a vortex,  antidot or tri-junction by tunneling into
it from a normal metal.  A signature of the zero mode would
be a zero bias anomaly, which would have a characteristic
current-voltage relation \cite{bolech07,law09}.

Another venue for Majorana fermions on a topological insulator surface is a
linear interface between superconducting and magnetically gapped regions
\cite{fukane08,tanaka09,linder10}.  This
leads to a 1D chiral Majorana mode, analogous to the edge state of
a 2D topological superconductor (Fig. \ref{fig:scedge}(e)).  This can
be used to construct a novel interferometer for Majorana
fermions \cite{fukane09b,akhmerov09}.  Fig. \ref{fig:majorana}(c) shows a
superconducting island surrounded by magnetic regions with a magnetic
domain wall.  The chiral Dirac fermions on the magnetic domain wall
incident from the left split into two chiral Majorana fermions on opposite sides of the
superconductor and then recombine.  If the superconductor encloses a
flux $\Phi=h/2e$, then the Majorana fermions pick up a relative minus
sign - analogous to the Aharonov Bohm effect.  This has the effect of
converting an incident electron into an outgoing hole, with a Cooper
pair of electrons absorbed by the superconductor.  This could be
observed in a three terminal transport setup.

Majorana bound states can also be engineered at the edge of a 2D quantum spin Hall insulator
utilizing magnetic and superconducting energy gaps
(Fig \ref{fig:majorana}(d)) \cite{fukane09a,nilsson08}.  This and other geometries
can in principle be used to test the inherent non-locality of Majorana fermion
states \cite{fu10}

\section{Conclusion and Outlook}
\label{sec:outlook}

Though the basic properties of topological insulators have been
established, the field is at an early stage
in its development.  There is much work to be done to
realize the potential of these new and fascinating materials.
In this concluding section we will discuss some very recent
developments and look toward the future.

In the history of condensed matter physics, the single most important
ingredient in the emergence of a new field is the perfection of the
techniques for producing high quality materials.  For example, the
intricate physics of the fractional quantum Hall effect would never
have emerged without ultra high mobility GaAs. Topological insulator
materials need to be perfected, so that they actually insulate. There
has been substantial progress in this direction.  For instance,
transport experiments on Ca doped crystals of Bi$_2$Se$_3$ show clear
insulating behavior below around 100$^\circ$K \cite{checkelsky09}.
However, the electrical resistance saturates at low temperature,
and the surface currents appear to be overwhelmed by either bulk
currents or currents in a layer near the surface.  This is
a challenging problem because narrow gap semiconductors are very
sensitive to doping. Nonetheless, it seems clear that there is ample
room for improvement. Thin films produced, for instance, by
mechanical exfoliation (as in graphene), or catalytically generated
Bi$_2$Se$_3$ ribbons and wires \cite{peng09} may be helpful in this
regard.  A particularly promising new direction is the growth of
epitaxial films of Bi$_2$Se$_3$
\cite{zhangg09,zhangy10} and Bi$_2$Te$_3$ \cite{li10}. This has led to the recent observation of Landau
quantization in the Dirac surface states \cite{cheng10,hanaguri10}. It will be interesting
to observe such quantization in a transport study.  Present materials pose
a challenge due to competition between surface and bulk states \cite{taskin09,checkelsky09}.
The detailed
study of the electronic and spin transport properties of the surface
states is called for, complemented by a host of other probes, ranging
from optics to tunneling spectroscopy.

Another direction for future innovation will be the study of {\it
heterostructures} involving topological insulators and other
materials.  In addition to providing a means for protecting and controlling the
population of the surface states, such structures could provide a step towards the
longer term goal of engineering exotic states, such as Majorana
fermions, with the surface states.  There are many materials problems
to be solved in order to find appropriate magnetic and
superconducting materials which exhibit the appropriate proximity
effects with the surface states, and detailed experiments will be
necessary to characterize those states.  An exciting recent development
along these lines is the discovery of superconductivity in Cu doped
Bi$_2$Se$_3$ \cite{hor10a}.  One can imagine devices fabricated
with techniques of {\it modulation doping} that have proved extremely
powerful in semiconductor physics.  One can also imagine other devices that
integrate magnetic materials with topological insulators, to take advantage of
the special spin properties of the surface states.

Topological insulating behavior is likely to arise in other classes
of materials, in addition to the binary compounds Bi$_{1-x}$Sb$_x$,
Bi$_2$Te$_3$, Bi$_2$Se$_3$ and Sb$_2$Te$_3$.
If one expands ones horizon to ternary compounds (or beyond) the
possibilities for exotic materials multiply.  Candidate materials will be narrow
gap semiconductors which include heavy elements.  One intriguing class
of materials are transition metal oxides involving iridium.
\textcite{shitade09} have predicted that
Na$_2$IrO$_3$ is a weak topological insulator.  \textcite{pesin10}
have suggested that certain iridium based pyrochlore compounds
may be strong topological insulators.  Since these materials
involve $d$ electrons, a crucial issue will be to understand the interplay
between strong electron-electron interactions and the spin-orbit interaction.
Very recent theoretical work predicts topological insulator behavior
in ternary Heusler compounds\cite{lin10a,chadov10} and other materials
\cite{lin10b,lin10c,yan10}.
These are exciting new directions where further theoretical and experimental
work is called for.

Topological superconductors present another exciting frontier direction.
In addition to observing the surface Majorana modes predicted for $^3$He B \cite{chung09}, it will be very
interesting to predict and observe electronic topological superconductors,
to characterize their surface modes
and to explore their potential utility.  To this end there
has been recent progress in developing methods to theoretically identify topological
superconductors based on their band structure \cite{qi10,fuberg10}.

More generally there may be other interesting correlated states related to topological
insulators and superconductors.  For example, \textcite{levin09} showed that
a ${\cal T}$ invariant {\it fractional} quantum spin Hall state can be topologically stable.
This points to a general theoretical problem associated with the
topological classification of {\it interacting} systems: How do we
unify the seemingly different notions of topological order
epitomized by \textcite{thouless82} and by \textcite{wen95}?
The topological field theories studied by
\textcite{qihugheszhang08}, as well as tantalizing connections
with string theory \cite{ryu10} provide steps in this direction, but a complete
theory remains to be developed.

To conclude, the recent advances in the physics of topological insulators have been driven
by a rich interplay between theoretical insight and experimental
discoveries.  There is reason for optimism that this field
will continue to develop in exciting new directions.

\section*{Acknowledgments} C.L.K. thanks E. J. Mele, L. Fu and
J. C. Y. Teo for collaboration and NSF Grant DMR 0906175 for support.
M.Z.H. thanks L. A. Wray for assistance preparing figures and
A. Bansil, R. J. Cava, J. G. Checkelsky, J. H. Dil, A.
V. Fedorov, Y. S. Hor, D. Hsieh, H. Lin, F. Meier, N. P. Ong, J.
Osterwalder, L. Patthey, D. Qian, P. Roushan, L. A. Wray, Y. Xia and A. Yazdani  for
collaborations and U.S. DOE (Grants No. DE-FG-02-05ER46200, No. AC03-76SF00098, and No. DE-FG02-
07ER46352), NSF [Grant No. DMR-0819860 (materials)
DMR-1006492], Alfred P. Sloan Foundation, and Princeton
University for support.

\bibliographystyle{apsrmp}

\end{document}